\documentclass[nofootinbib,floats,letterpaper,floatfix,eqsecnum,prd,aps,10pt]{revtex4}

\usepackage{dcolumn,epsfig,minitoc}
\usepackage{hyperref}
\usepackage{amssymb,amsmath}
\usepackage[usenames]{color}

\def\nn{\nonumber}

\def\be{\begin{equation}}
\def\ee{\end{equation}}
\def\beq{\begin{eqnarray}}
\def\eeq{\end{eqnarray}}

\def\IL{\relax{\rm I\kern-.18em L}}

\def\nn{\nonumber}

\def\m{\mu}
\def\n{\nu}
\def\r{\rho}

\def\b{\beta}

\def\l{\lambda}

\def\D{\Delta}
\def\w{\omega}

\def\nb{\nonumber}

\def\ba{\begin{eqnarray}}
\def\ea{\end{eqnarray}}

\begin{document}

\title{Holography of charged dilatonic  black branes at finite
temperature}

 \author{Mariano Cadoni}\email{email: mariano.cadoni@ca.infn.it }
\affiliation{Dipartimento di Fisica, Universit\`a di Cagliari and
INFN, Sezione di Cagliari - Cittadella Universitaria,
09042 Monserrato, Italy. }
\author{Paolo Pani}\email{email: paolo.pani@ist.utl.pt}
\affiliation{CENTRA, Departamento de F\'{\i}sica, 
Instituto Superior T\'ecnico, Universidade T\'ecnica de Lisboa - UTL,
Av.~Rovisco Pais 1, 1049 Lisboa, Portugal.}


\date{\today}

\begin{abstract}
We investigate  bulk and holographic features of  
finite-temperature  black brane 
solutions of 4D anti-de Sitter Einstein-Maxwell-dilaton-gravity
(EMDG). We
construct, 
numerically, black branes endowed with non trivial scalar 
hairs for broad classes of EMDG. We consider both exponential and
power-law forms for 
the coupling functions, as well as several charge configurations: 
purely electric, purely magnetic and dyonic solutions. At finite
temperature the
field theory
holographically 
dual to these black brane solutions has a rich and interesting 
phenomenology reminiscent of
electron motion in
 metals:  phase transitions triggered by nonvanishing VEV
of 
scalar operators,  non-monotonic behavior of the electric 
conductivities as function of the frequency and of the 
temperature, Hall effect and sharp synchrotron resonances of the
conductivity in presence       
of a magnetic field. Conversely, in the zero temperature limit the 
conductivities for these models show a universal   behavior.
The optical conductivity has a power-law behavior
as a function of  the frequency, whereas the DC conductivity is 
suppressed at small temperatures. 

\end{abstract}


\maketitle

\setcounter{tocdepth}{1}
\tableofcontents


\section{Introduction}
One of the most productive applications of the AdS/CFT
\cite{Maldacena:1997re} 
correspondence is obtained when one considers the regime in which
the classical gravity approximation is reliable. 
In this regime we can  deal with strongly coupled quantum field
theories (QFTs) in 
$d-1$ dimensions   by  investigating  classical gravity
in $d$ dimensions.   
In particular, this holographic approach has been recently used to 
provide  techniques
for the computation of  thermodynamical and transport properties
of strongly interacting quantum field theories, which could be 
relevant in the description of condensed matter phenomena 
\cite{Sachdev:2008ba, Herzog:2009xv, Hartnoll:2009sz,
McGreevy:2009xe}.

The basic structure of the holographic correspondence is given by a 
black hole (black brane)
in 
the $d$-dimensional bulk, which is holographically dual to a
thermal QFT in 
$d-1$ dimensions. This framework has
been used in 
several cases to investigate the hydrodynamical limit 
of the QFT,  to compute 
spectral functions and so on.
However, the most interesting results come up when we 
consider a thermal QFT with finite charge density, i.e a charged black
hole in 
the bulk.
An even richer structure is obtained if we include in the bulk  a
charged scalar field with a minimal (covariant) coupling to the gauge
potential.
Below a critical temperature the bulk theory allows for
solutions with a non-trivial profile of 
the scalar field. This   corresponds 
to the formation of a charged
condensate in the dual theory that breaks spontaneously a global
$U(1)$ symmetry. The new phase is therefore characterized by
phenomena typical of superfluid or superconducting systems, as one 
can show by using the basic rules of the AdS/CFT correspondence, 
i.e by studying  the response of  small perturbations of the black
hole background \cite{Gubser:2008wv,
Hartnoll:2008kx,Hartnoll:2008vx}.

This general idea of holographic superconductor has generated in the
last 
couple of years a flurry of 
activity on AdS gravity whose holographic QFT duals have transport 
features with a metal or metal-like behavior \cite{Rey:2008zz,
Lee:2008xf,Liu:2009dm,Cubrovic:2009ye,Faulkner:2009wj,McInnes:2009ux,
Faulkner:2010gj,Lugo:2010qq,D'Hoker:2010rz}.

Most of the efforts for understanding the holography of charged black 
branes  have focused on the case of  AdS 
Einstein-Maxwell
gravity with a charged scalar fields
minimally coupled to the electromagnetic field. 
However, there are several reasons for extending the investigation to 
the case in which the scalar is non-minimally coupled  to the $U(1)$ 
field, the so called Einstein-Maxwell-dilaton gravity (EMDG): 
$1)$ Non-minimal
couplings of the form $f(\phi) F^{2}$ between a scalar fields $\phi$
and the Maxwell tensor  are very common in supergravity and in the
low-energy effective action of string theory models. 
$2)$ Exact, charged
dilaton black hole solutions with AdS asymptotics are known in some 
cases, for instance
the family of four-charge black holes in ${\cal N}=8$
four-dimensional gauged supergravity \cite{Duff:1999gh}.
$3)$ Charged black brane solutions  with AdS asymptotics of EMDG have
a 
rather interesting thermodynamical phase structure 
\cite{Cadoni:2009xm,Charmousis:2010zz,Doneva:2010iq}.
For instance, when the coupling function $f$ satisfies the requirement
$f'(0)=0$ the model allows for a phase transition between  the 
Reissner-Nordstrom (RN) black brane   and  a charged black brane  
endowed with 
a scalar hair \cite{Cadoni:2009xm}, corresponding in the dual field 
theory to the formation of a scalar condensate.
$4)$ In the few cases in which EMDG has been investigated within
the 
holographic perspective, the models have shown a rather rich and 
interesting phenomenology   
\cite{Gubser:2009qt,Goldstein:2009cv,Cadoni:2009xm,Goldstein:2010aw,
Chen:2010kn,Lee:2010qs,Lee:2010ii,Liu:2010ka,Lee:2010ez}.
$5)$ Finally, the black brane solutions of EMDG are good candidates
for 
gravitational backgrounds holographically dual to Lifshitz-like 
theories
\cite{Goldstein:2009cv,Bertoldi:2011zr,Bertoldi:2010ca,Perlmutter:2010qu}.

Until now  
investigations of charged dilaton AdS black branes and their 
holographic features  have considered almost exclusively  the cases 
of coupling functions $f(\phi)$ with exponential behavior in the zero 
temperature limit  
\cite{Goldstein:2009cv,Charmousis:2010zz,Perlmutter:2010qu}.
In this paper  we  extend   the investigation of
the
black brane solutions of EMDG in four dimensions (4D) and of their
holographic
properties.
On the one hand we  complete the investigation of EMDG with  
exponential coupling $ f\sim e^{\alpha \phi}$, by considering 
electrically charged  solutions of the model at finite, nonvanishing, 
temperature. On the other hand, we extend the investigation to 
electrically charged black brane solutions of models 
with a power-law coupling function $f\sim \phi^{m}$ both
at 
finite temperature and in the zero temperature limit.
Last but not least, we consider  dyonic 
solutions of
models characterized by $f'(0)=0$, thus allowing a phase transition 
between the dyonic RN solution and a dyonic black brane with a
nontrivial scalar
hair.

After constructing, 
numerically, the black brane solutions endowed with  scalar 
hairs for our models, we investigate, using the AdS/CFT
correspondence,
the
field theory
holographically 
dual to these black brane solutions both at finite temperature and in
the 
zero temperature limit.
We show that at finite temperature the dual field theory 
presents a  rich 
phenomenology: phase transitions triggered by nonvanishing VEV of 
scalar operators,  non-monotonic behavior of the electric 
conductivities, Hall effect and sharp synchrotron resonances of the
conductivity in presence       
of a magnetic field. 
On the other hand, in the zero temperature limit
the 
optical conductivity for these models shows a universal  power-law
behavior
as a function 
of  the frequency, whereas the DC conductivity is suppressed at small 
temperature  and in general scales as $T^{2}$.

Our investigations suggest an intriguing holographic picture for the 
black brane solutions of EMDG. We have an interpolation between some 
features of electron motion in metals at finite temperature and an 
universal strong repulsion behavior, characteristic of charged 
plasmas, at zero temperature.

The plan of the work is the following. In Sect.~\ref{sec:eqs} we
present general models of EMDG and the
equations of motion. Sect.~\ref{sec:CDBBs} is devoted to the study of
electrically charged dilatonic black branes in these models and their
holographic properties at finite temperature. The zero temperature
limit is discussed in Sect.~\ref{sect:zerot}. In
Sect.~\ref{sec:DDBBs} we discuss dyonic dilatonic black branes at
finite temperature, while in Appendix~\ref{appendix} we briefly
present purely magnetic solutions. Finally, we draw our conclusion in
Sect.~\ref{conclusion}.
\section{ Einstein-Maxwell-dilaton gravity in Anti
de Sitter spacetime}\label{sec:eqs}
In this paper we will consider  general models of
EMDG in 4D, which is described by the action
\be S=\int d^4x\sqrt{-G}\,{\cal L}=\int d^4
x\sqrt{-G}\left(R-\frac{f(\phi)
}{4}F^2-\frac{1}{2}\partial^\mu\phi\partial_\mu\phi-V(\phi)\right)\,,
\label{lagr_genB}
\ee
where $\phi$ is a scalar field (the dilaton) and $F$ is the field 
strength for the Maxwell field. 
The model is parametrized by two functions: the coupling $f(\phi)$ 
between the 
scalar   and the Maxwell tensor and the potential  $V(\phi)$
describing 
the self-interaction of the scalar. 
Non-minimal
couplings, $f(\phi) F^{2}$ between a scalar fields $\phi$
and the Maxwell tensor  emerge naturally in supergravity and in the
low-energy effective action of string theory.

In asymptotically flat spacetime, exact, charged black holes (or
black branes) solutions 
of EMDG carrying non trivial scalar
hairs (i.e. 
different from the usual Reissner-Nordstrom (RN) solutions) 
are known since a long  time
\cite{Gibbons:1987ps,Garfinkle:1990qj,Cadoni:1993yt,Monni:1995vu,Charmousis:2009xr}.
These solutions
involve a non-constant scalar field and differ  from
the RN black hole both with respect to the causal structure of the 
spacetime and to the thermodynamical behavior.
Conversely, for charged
dilaton black holes with AdS asymptotics  exact solutions with scalar 
hairs   are known 
only in few cases~\cite{Duff:1999gh,Mignemi:2009ui}, the most 
important one  being represented  by
the family of four-charge black holes in ${\cal N}=8$
four-dimensional gauged supergravity~\cite{Duff:1999gh}.

In recent years there has been a renewed interest for black hole 
solutions of EMDG with AdS 
asymptotics. This 
interest has been triggered by the 
gauge/gravity duality~\cite{Maldacena:1997re,Horowitz:2006ct} 
and  by the search for gravitational duals of 
strongly coupled condensed matter systems with finite charge 
density~\cite{Sachdev:2008ba,Herzog:2009xv,Hartnoll:2009sz}.

However, until now the 
investigation of charged dilaton AdS black holes and their 
holographic features  has been  almost completely restricted to case 
of coupling functions $f(\phi)$ with exponential behavior
\cite{Goldstein:2009cv,Cadoni:2009xm,Charmousis:2010zz} 
and to electric charged solutions (see however
Ref.~\cite{Goldstein:2010aw}).
There are several reasons behind this choice. Exponential coupling 
functions $f(\phi)$  are rather natural in low-energy effective 
string theory. Furthermore, it is  very difficult to find exact 
solutions of  generic AdS Maxwell-dilaton gravity and typically one
has to resort to numerical calculation. Conversely, an exponential
form for
$f(\phi)$ enables one   
to find exact solutions at least in the extremal limit, where they
take a 
Lifshitz-like form \cite{Goldstein:2009cv}. 

Although the  pure exponential form 
$f(\phi)= e^{\alpha \phi}$ does not allow for a RN  solution,  
simple deformations  preserving an exponential behavior on the 
horizon, such as $f(\phi)= \cosh{\alpha \phi}$, do allow for it.
The coexistence in the model of the RN solution together with
solutions with 
scalar hairs is crucial for having a phase
transition~\cite{Cadoni:2009xm}.

In this paper we will not limit ourself to the case of exponential 
coupling functions $f(\phi)$ but we will extend  our investigation to
the
black brane solutions in a broad class of EMDG models.

The equations of motion stemming from the action (\ref{lagr_genB})
read
\beq\label{max_scal_eq} &&\nabla_\mu\left(f(\phi)F^{\mu\nu}\right)=0
\,, \nb \\
&&\nabla^2\phi=\frac{dV(\phi)}{d\phi}+\frac{df(\phi)}{d\phi}\frac{F^2}{4}\,,
\\
&&R_{\mu\nu} - \frac{1}{2} G_{\mu \nu} R  = 
\frac{f(\phi)}{2}\left( F_{\mu \rho} F_{\nu}^{\rho} - \frac{G_{\mu
\nu}}{4} F^{\rho\sigma} F_{\rho\sigma}\right)  + \frac{1}{2} \left (
\partial_{\mu}\phi\partial_\nu \phi - \frac{G_{\m\n}}{2}
\partial^{\rho}
\phi\partial_{\rho}\phi \right ) - \frac{G_{\m\n}}{2} V(\phi) \,. \nb
\label{einsteineq} \eeq

We will look for static solutions of the previous equations
with translational symmetry in two spatial directions (black branes)
carrying electric and/or magnetic charges.
The metric and the scalar field  have the form
 \be
ds^2=-g(r)e^{-\chi(r)}dt^2+\frac{dr^2}{g(r)}+r^2(dx^2+dy^2)\,, \qquad
\phi=\phi(r).
\label{metric_ansatz_BR}
\ee
Since we are interested in the holographic features of our models, 
we will only  constrain the form 
of the coupling functions $f(\phi)$ and $V(\phi)$ by imposing
conditions at 
$r=\infty$ (corresponding to the UV region of the dual
field theory) and at 
the horizon of the extremal, zero temperature, solution
(corresponding 
to the IR region of the  dual field theory).
 
We first require the solutions to  be 
asymptotically AdS and  the potential $V(\phi)$
to allows for stable AdS vacua. 
Assuming for simplicity that $V(\phi)$ has   only
one extremum, we can consider without loss of generality 
that $\phi=0$ as $r\to \infty$. 
The potential can be now expanded for small
values of the field as\footnote{The easiest way to include a massive
scalar field is by considering $V(\phi)=-{6}/{L^2}+{\b}/(2
L^2)\phi^2$. Although the numerical results presented in this paper
where obtained
using this form of $V(\phi)$, other choices give qualitatively
similar results, provided they satisfy Eq.~\eqref{Vexp}. We
explicitly checked that the potential
$V(\phi)=-{6}/{L^2}\cosh(\phi/\sqrt{3})$ gives the same qualitative
results as the polynomial form with $\beta=-2$.}
 \be V(\phi)=-\frac{6}{L^2}+\frac{\b}{2
L^2}\phi^2+{\cal O}(\phi^3)\,, \label{Vexp}\ee 
where $L$ is the AdS radius and
$\b$ parametrizes the mass of the scalar field, $m_{s}^2 L^2 = \b$.
The
AdS
vacuum is stable if the mass parameter satisfies the
Breitenlohner-Freedman (BF) bound $\b \ge - 9/4$
\cite{Breitenlohner:1982bm}.
On the other hand the  coupling functions $f(\phi)$ is only
constrained  by
$f(0)=1$ \footnote{ In general, we should only require 
$f(0)$ to be finite, however $f(0)$ can be set to the unit by a 
redefinition of the charges.}.  
Notice that in general we will not require the AdS-RN black hole  to
be a  solution
of the equations (\ref{max_scal_eq}). This would imply the additional 
condition $f'(0) = 0$.

Secondly, in the extremal limit (the zero temperature solution) 
we can assume without loss of generality that the horizon is located 
at $r=0$. The IR behavior of the field theory dual to our dilaton
gravity 
model will be determined by the  asymptotic 
expansion of the solution near $r=0$. Because in the near-horizon 
region we expect this dual theory to be strongly coupled, we  assume
that 
$\phi\to\infty$ as $r\to 0$ and $f(\infty)=\infty$. The asymptotic
form 
of the functions $f(\phi)$ 
and $V(\phi)$ is therefore determined by the leading term in the
$\phi\to\infty$
expansion of the field.

The class of functions $f(\phi)$ and $V(\phi)$ satisfying the
requirements above
is rather broad. However, we will not expect the qualitative behavior
of the dual field  theory to depend strongly on the details of the 
functional form of $f(\phi)$ and $V(\phi)$. The essential
information is contained,  for what concerns the  UV region,
in  $V''(0)$, $f'(0)$ and $f''(0)$ (primes denote derivative with
respect to $\phi$). On the other hand the 
form of the potential $V(\phi)$  seems to be quantitatively  but not 
qualitatively relevant for the description of the zero temperature 
extremal limit \cite{Goldstein:2009cv,Cadoni:2009xm}. Thus, the
behavior in the IR
region will be determined by the leading terms in the $\phi\to\infty$ 
expansion of $f(\phi)$. Obviously, the most interesting cases are 
represented by an exponential $f\sim e^{\alpha \phi}$ and  
a power-law $f\sim \phi^{m}$ behavior.

A vanishing value of 
$f'(0)$ discriminates between models for which a phase transition 
between the AdS-RN black brane and a solution with scalar hairs is 
possible. The values of $V''(0)$, and $f''(0)$  determine both the 
parameter region where the phase transition takes effectively place 
and the specific  behavior of the transport coefficients of the dual 
theory \cite{Cadoni:2009xm}.

It follows that a rough classification of the Maxwell-dilaton-gravity
model  
with interesting holographic features  can be simply given in terms
of a 
vanishing/nonvanishing $f'(0)$ and by  the leading terms in 
the $\phi\to\infty$ expansion of $f(\phi)$.  

Electric charged solutions for models with $f'(0)=0$, both at finite
and zero
temperature,
has been investigated  in Ref. \cite{Cadoni:2009xm}. 
Models with exponential coupling function $f(\phi)$   have been 
investigated in the zero temperature limit 
$T=0$  both in the case of electric  and dyonic solutions 
\cite{Goldstein:2009cv,Goldstein:2010aw}.

In this paper we complete the investigation of the holographic 
properties  of the most interesting EMDG 
models, by considering:\\ 
$a)$ Electrically charged  solutions of models
with  
$ f\sim e^{\alpha \phi}$ at 
finite temperature ;\\
$b)$ Electrically charged solutions  of models
with  
$f\sim \phi^{m}$ ($m\neq 1$) both at 
finite temperature and in the zero temperature limit;\\ 
$c)$ Electrically charged solutions  of models
with  
$f\sim \phi$ both at 
finite temperature and in the zero temperature limit\\
$d)$ Dyonic 
solutions of
models with $f'(0)=0$.

Notice that we consider separately the power-law case and the linear 
case. This is because  the former case allows for a phase  
transition to the AdS-RN solution   ($f'(0)=0$), whereas the latter 
does not  ($f'(0)\neq 0$).

\section{Electrically charged dilatonic black branes at finite
temperature}\label{sec:CDBBs}
In this section we derive, numerically,  electrically charged black 
brane solutions of the 
EMDG \eqref{lagr_genB} at finite 
temperature with
exponential,  power-law and linear coupling and investigate their 
holographic features.

Since only the $A_0$ component of
the gauge potential
is non-vanishing and $\phi=\phi(r)$, the equations
of motion  (\ref{max_scal_eq}) become 
\beq
\phi''+\left(\frac{g'}{g}-\frac{\chi'}{2}+\frac{2}{r}\right)\phi'(r)-\frac{1}{g}\frac{d
V}{d \phi}+\frac{{A_0'}^2e^\chi}{2g} \frac{d f}{d
\phi}&=&0\,,\nonumber\\
(r^2 e^{\frac{\chi}{2}} f(\phi) A_0')'  &=& 0
\,,\nonumber\\
\chi'+\frac{r{\phi'}^2}{2}&=&0 \nonumber\,,\\
\frac{{\phi'}^2}{4}+\frac{{A_0'}^2e^\chi
f(\phi)}{4g}+\frac{g'}{rg}+\frac{1}{r^2}+\frac{V(\phi)}{2g}&=&0\, ,
\label{eq:BR_einstein} \eeq 
where now the prime  denotes a
derivative with respect to $r$. Charged dilatonic black branes
(CDBBs) solutions  of the field 
equations (\ref{eq:BR_einstein}) at finite temperature can be
computed numerically by using the method discussed in
Ref.~\cite{Cadoni:2009xm}
with slight modifications due to the form of $f(\phi)$.
\subsection{Numerical solutions}
The numerical procedure for solving the
field equations~\eqref{eq:BR_einstein} consists essentially in three 
steps (see Ref.~\cite{Cadoni:2009xm} for details).  First, we
characterize the solutions in terms of the  behavior of
the
fields
near the AdS boundary
at $r =\infty$ and near the horizon at $r =
r_h$.  Near the AdS boundary  the solutions are  specified
by four parameters: the chemical potential $\mu$, the charge
density  
$\r$ appearing in the expansion of the gauge potential $A_{0}= 
\mu-\r/r$, the black brane mass $M$ and the expectation values for
the 
operators dual to the scalar field ${\cal O}_-$, ${\cal 
O}_{+}$. These are determined 
in terms of the asymptotic expansion of the scalar 
field\footnote{In order to describe states of the dual field theory
with a non
vanishing expectation value for the operator dual to the scalar
field, the asymptotic expansion $(\ref{bc_scalarB})$ should contain
only normalizable modes~\cite{Klebanov:1999tb}. When
$m^2 L^2 \ge -5/4$ this requires the boundary condition ${\cal O}_-
= 0$, whereas, when $- 9/4 < m^2 L^2 < -5/4$ two distinct choices
are possible~\cite{Hertog:2004rz}: ${\cal O}_- = 0$ or  ${\cal
O}_{+} = 0$. Finally, when the Breitenlohner-Freedman (BF)
bound~\cite{Breitenlohner:1982bm} is saturated, $m^2 L^2=-9/4$, the
asymptotic behavior has a logarithmic branch, whose coefficient is
required to vanish.}:

\beq 
\phi \sim \frac{{\cal O}_-}{r^{\D_-}}+\frac{{\cal
O}_{+}}{r^{\D_+}}\,,\qquad \D_{\pm}=\frac{3\pm\sqrt{9+4\beta}}{2}\,.
\label{bc_scalarB} 
\eeq
Near the horizon  the solutions
are completely specified by four parameters: 
the horizon radius $r_h$, $A_0'(r_h)$, $\chi(r_h) \equiv \chi_h$ and
$\phi(r_h) \equiv \phi_h$. The  black
brane temperature $T$  can be expressed in terms of these parameters.
Secondly, we reduce the numbers of parameters exploiting 
the scaling symmetries of
the equations of motion. Using these symmetries  the number 
of independent parameters specifying the solutions can be effectively
reduced 
to two. Furthermore, we choose one of
the independent parameters at the horizon such that
either ${\cal O}_-=0$ or ${\cal O}_+=0$. This is enough to describe
the solution in terms of a single parameter. 

Finally, we integrate numerically the equations of motion from the
horizon to the AdS asymptotics, using a shooting method to impose
either ${\cal O}_-=0$ or ${\cal O}_+=0$. The result of this
integration is a one-parameter family of solutions, the parameter
being the black brane temperature.

We shall directly focus on the results of the numerical integration
for 
the various models under consideration, referring  
to Ref.~\cite{Cadoni:2009xm}
for further details on the numerical technique used.

\subsubsection{Exponential coupling function}
\label{sect::expocou}
Charged black branes solutions of dilaton gravity with an exponential 
coupling function at $T\sim0$ have been recently studied in great
detail in Refs.~\cite{Goldstein:2009cv,Goldstein:2010aw}. Here we
consider  the solution at finite temperature and 
focus on models defined by

\beq\label{f51}
f(\phi)=e^{\alpha\phi}\,,\qquad
V(\phi)=-\frac{6}{L^2}+\frac{\b}{2L^2}\phi^2\,.\label{expf}
\eeq

The model discussed in Ref.~\cite{Goldstein:2009cv} is obtained for
$\beta=0$. We are interested in finite temperature effects for a
generic mass of the scalar field. 

Models with coupling functions given by Eqs.~\eqref{expf} do not allow
for 
AdS-RN black branes, which may be  regarded has
 a pleasant feature, since the large entropy of extremal AdS-RN
black holes poses some problems of interpretation from the
holographic perspective. Conversely, CCBBs have a vanishing entropy
in the extremal limit. Moreover, they exist at any temperature, since
no phase transition occurs in these models. 

Our method provides
numerical
solutions for CDBBs at finite temperature. They describe hairy black 
branes, i.e. black branes endowed with a non trivial configuration
for 
the scalar field $\phi$. These hairy solutions 
can be completely characterized in terms of the expectation value of
one of the neutral
boundary
operators  ${\cal O}_{\pm}$ (the scalar condensate) defined in Eq.
\eqref{bc_scalarB}
as a function of the black hole  temperature $T$. 

\begin{figure}[ht]
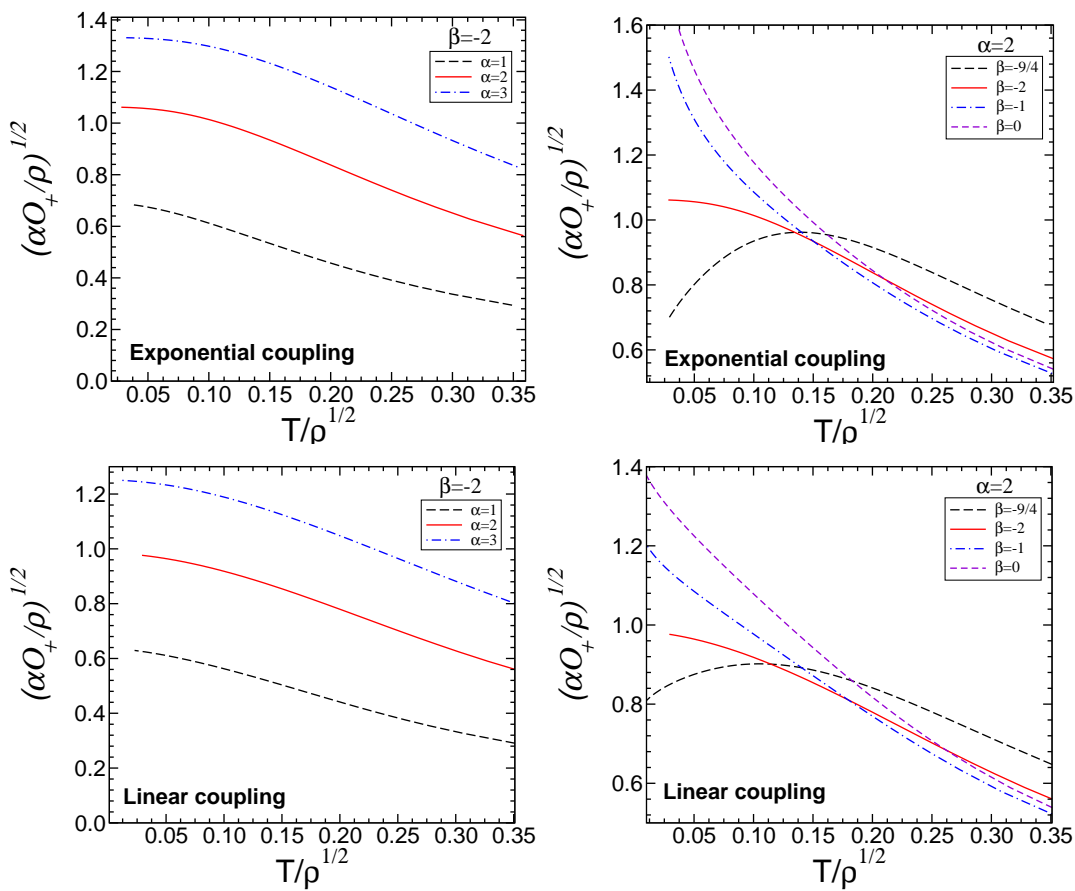

\begin{center}
\begin{tabular}{cc}
\epsfig{file=Plots/CDBB_O2_a.eps,width=7cm,angle=0}&
\epsfig{file=Plots/CDBB_O2_b.eps,width=7cm,angle=0}\\
\epsfig{file=Plots/CDBB_O2_a_lin.eps,width=7cm,angle=0}&
\epsfig{file=Plots/CDBB_O2_b_lin.eps,width=7cm,angle=0}
\end{tabular}
\caption{The scalar operator ${\cal O}_+$ as
a function of the
temperature for several values of $\alpha$ when $\beta=-2$ (left
panels) and several values of $\beta$ when $\alpha=2$ (right panels).
In the upper panels and lower panels we have used models with
$f(\phi)=e^{\alpha\phi}$ and with linear coupling
$f(\phi)=1+\alpha\phi$, respectively. Results are qualitatively
similar in the two cases.
\label{fig:CDBBs_operators}}
\end{center}
\end{figure}

The behavior of some of these boundary operators as function of $T$
are shown in the upper panels of Fig.~\ref{fig:CDBBs_operators} for
several values of $\alpha$ and
$\beta$. 
The dependence of the condensate from $\alpha$ is quite simple.
Roughly speaking, larger values of $\alpha$ shift the condensate up.
However, the dependence on the  mass $m_s$ of the scalar  is more
involved. In fact,
the condensate monotonically depends on the temperature for
$\beta=0,-1,-2$, but it is a non-monotonic function in the BF limit,
$\beta=-9/4$. Notice that the condensate exists at any temperature
and, although not shown in Fig.~\ref{fig:CDBBs_operators}, it
vanishes asymptotically when $T\gg \sqrt{\rho}$.

\subsubsection{Linear coupling function}

Let us now  consider a linear  coupling function $f(\phi)$  and 
the usual self-interaction potential
\be\label{f6}
f(\phi)=1+\alpha\phi\,, \qquad
V(\phi)=-\frac{6}{L^2}+\frac{\beta}{L^2}\phi^2\,.
\ee
Similarly to the exponential case discussed in section 
(\ref{sect::expocou}), also here we have $f'(0)\neq 0$, forbidding 
the existence of AdS-RN solutions. Thus only hairy black hole
solutions 
characterized by a nonvanishing scalar condensate are allowed.
Because the coupling function $f(\phi)$ given by Eq.~(\ref{f6}) is
just the 
linear approximation near $\phi=0$  of the exponential coupling 
(\ref{f51}),  these solutions have the same $r\to \infty$ behavior as 
those discussed in section 
(\ref{sect::expocou}). Obviously the two sets of solutions differ in 
the near-horizon region.  
The behavior of the scalar condensate is shown in the lower panels of
Fig.~\ref{fig:CDBBs_operators} and it  
behaves in a way which is qualitatively 
similar to that described in the upper panels for exponential
couplings. This suggests that the salient properties of these
solutions are simply captured by the condition $f'(0)\neq0$.

\subsubsection{Power-law coupling function}\label{power-law}
\begin{center}
\begin{figure*}[ht]
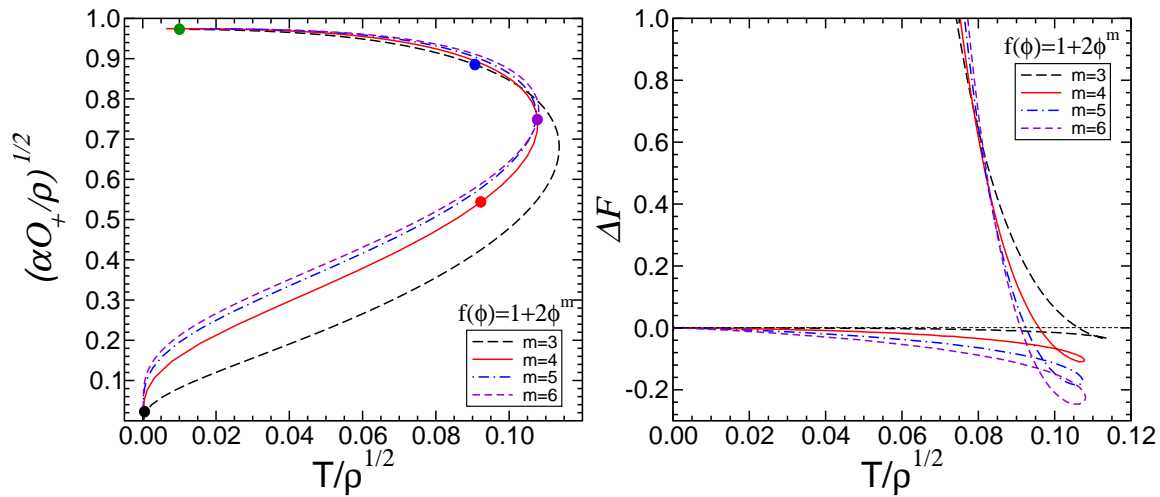

\begin{center}
\begin{tabular}{cc}
\includegraphics[scale=0.31,clip=true]{./Plots/PL_O2.eps}&
\includegraphics[scale=0.31,clip=true]{./Plots/PL_DF.eps}
\end{tabular}
\end{center}
\caption{Left panel: scalar condensate for models with a power-law 
coupling function $f(\phi)=1+\alpha\phi^m\,$ as a function of the
temperature. Markers correspond to the temperatures considered in
Fig.~\ref{fig:PL_cond2} (see Sect.~\ref{sect:zerot}). Right
panel: Difference in free energy  between  the AdS-RN black brane
and the hairy
black brane, $\Delta F=F_{RNBB}-F_{HBB}$ as a function of the
temperature. When $\Delta F>0$ the dilatonic solution is
energetically favored. We have considered  $\alpha=2$ and
$\beta=-2$.}
\label{fig:PL_O_F}
\end{figure*}
\end{center}

The method described above can be also used  to construct numerical 
solutions for the class of models  
with power-law coupling function,
\be
f(\phi)=1+\alpha\phi^m\,,\quad m > 1\,, \qquad
V(\phi)=-\frac{6}{L^2}+\frac{\beta}{L^2}\phi^2\,.\label{f_powerlaw}
\ee
The most important difference between this and the previously 
discussed cases is that $f'(0)=0$. This implies that in  these
models 
black brane solutions with non 
trivial scalar profile for the scalar field can coexist with the 
AdS-RN black brane for
any $m> 1$. In general, these  models will be therefore
characterized 
by a phase transition. Below a critical temperature $T_{c}$ the 
AdS-RN solution becomes unstable and  the  operator dual to the 
scalar field acquires a nonvanishing expectation value, i.e.  a
scalar condensate
forms.
This behavior is very similar to that observed in the class of 
models investigated in Ref.~\cite{Cadoni:2009xm}.
In the left panel of
Fig.~\ref{fig:PL_O_F} we show the expectation value for the scalar
operator ${\cal O}_+$ as a function of the temperature for different
values of $m$. Interestingly, near the critical temperature these
solution exist also for $T>T_c$ and, for a given temperature, there
exist
two branches of solutions. This is due to a non-monotonic behavior of
the function $T(\phi_h)$. 

Moreover, as shown in the right panel of Fig.~\ref{fig:PL_O_F}, in
the first branch the hairy dilatonic solutions have always  a larger
free energy than the corresponding AdS-RN black brane at the same
temperature. Hence, they are energetically less
favored and likely decay into AdS-RN black branes. On the other hand,
solutions in the second branch are unstable with respect to the
AdS-RN solution only at (relatively) high values of $T$. After the
turning point shown in the right panel of Fig.~\ref{fig:PL_O_F}, the
difference in the free energy $\Delta F$ changes sign and solutions
in the second branch are energetically favored all the way down to
the zero temperature limit (see Ref.~\cite{Aprile:2009ai} for similar
effects
in models of holographic superconductors with nonminimal couplings).
This behavior has to be compared with that 
arising in the EMDG models investigated 
in Ref. \cite{Cadoni:2009xm} (and also in the case of a scalar with 
covariant coupling with the gauge field  \cite{Gubser:2008wv,
Hartnoll:2008kx,Hartnoll:2008vx}). In these latter 
cases  the  hairy solution exists and, it is stable, only below the 
critical temperature. Above $T_{c}$ only the AdS-RN solution exists
\cite{Cadoni:2009xm,Gubser:2008wv,
Hartnoll:2008kx,Hartnoll:2008vx}.

As  we will discuss in detail in Sect.~\ref{sect:zerot},
in the zero temperature limit 
 the hairy  solutions are stable and show a behavior
which is largely independent from the parameters.
Indeed, the qualitative behavior does not depends on $m$, $\alpha$
and $\beta$. This universality can be traced back to the common
property
of these theories, which have $f'(0)=0$ regardless of the
values
of $m$ and $\alpha$.
%

\subsection{Holographic properties}
\label{holoprop}
Holographic properties of CDBBs with exponential coupling function 
$f\sim \exp(\alpha \phi)$
have been investigated in great
detail in Refs.~\cite{Goldstein:2009cv,Goldstein:2010aw}, in the zero
temperature limit (see also Ref.~\cite{Charmousis:2010zz}). In this
limit, an approximate method can be
developed in order to obtain analytical results for thermodynamical
properties and transport coefficients in the dual field theory. At
zero temperature and when $\beta=0$, some of these holographic 
features of
CDBBs are mostly universal, i.e. they do not depend on $\alpha$. The
dual theory is reminiscent of a charged plasma at
$T\sim0$~\cite{Goldstein:2010aw}. However, some transport properties,
such as the optical conductivity, can depend on the actual form of
the self-interaction potential $V(\phi)$~\cite{Cadoni:2009xm}, or on
the number
of spacetime dimensions~\cite{Chen:2010kn}, even in the zero
temperature limit. See also
Refs.~\cite{Charmousis:2009xr,Doneva:2010iq} for related studies.

Here, we extend  the investigation to  a large class of CDBBs 
and show that some holographic properties at
finite (but non-vanishing) temperature are not universal. Indeed,
qualitative differences arise depending on the particular model at
hand.
This is also true even for the
very same configuration explored in Ref.~\cite{Goldstein:2010aw},
i.e.  an exponential coupling function 
$f\sim \exp(\alpha \phi)$ and  $\beta=0$. In this case qualitative
differences arise depending on the value of $\alpha$.

The transport properties in the field theory dual to CDBBs  show a 
rather rich structure.
For instance, as we will see later on this paper, the electrical
conductivity 
as a
function of the temperature shows some remarkable and non-trivial
behavior, previously observed in Ref. \cite{Cadoni:2009xm} (see
also Ref.~\cite{Hartnoll:2009ns,Liu:2010ka} for similar results in
the context of strange metals and holographic superconductors,
respectively).

The AdS/CFT correspondence provides a
precise prescription for describing  transport phenomena of  the
field 
theory dual to CDBBs in terms of  perturbations of fields in the bulk.
This is  in particular true  for  the electrical conductivity, which 
can be derived from the
equations governing the fluctuations of the gauge field component
$A_x$ and of the metric component $G_{tx}$
\cite{Hartnoll:2009sz}.
In the case of a purely electric  background, perturbations of  $A_x$
with zero spatial momentum and
harmonic time dependence decouple from all the other modes.

The
perturbation $A_x$ is obtained by solving the
 equation \cite{Goldstein:2009cv,Cadoni:2009xm}
\be
A_x''+\left[\frac{g'}{g}-\frac{\chi'}{2}+\frac{f'(\phi)}{f(\phi)}\right]A_x'+
\left(\frac{\omega^2}{g^2}-\frac{{A_0'}^2 f(\phi)}{g}\right)
e^\chi
\, A_x=0\, , \label{eq:vp} \ee
with purely ingoing boundary conditions at the horizon.

The electric 
conductivity  of the dual field theory is given by
\cite{Hartnoll:2008vx} 
\be
\sigma=-i\frac{A_x^{(1)}}{\omega A_x^{(0)}}\,
\label{definition_sigma} \ee
where $A_x^{(0)}$ and $A_x^{(1)}$  are
defined by the asymptotic behavior of the fluctuation at infinity $
A_x\sim A_x^{(0)}+{A_x^{(1)}}/{r}.$

The electrical conductivity  $\sigma$, in particular its dependence 
from the frequency $\omega$, can be also calculated by recasting Eq. 
(\ref{eq:vp}) in the form of a Schr\"odinger-like  equation
\cite{Horowitz:2009ij}. 
 The conductivity can be expressed  in term of the  reflection
coefficient ${\cal 
 R}$  for a quantum particle incident from the right on a 
potential barrier, generated by an effective potential $V_{s}¥(z)$
 \cite{Horowitz:2009ij,Cadoni:2009xm}:

\be
\sigma(\omega)
=\frac{1-{\cal R}}{1+{\cal R}}-\frac{i}{2 \omega} \left [
\frac{1}{f}\frac{d f}{dz}\right ]_{z=0} 
\, . \label{sigma} \ee
where the coordinate   $z$ is defined by $dr/dz= g \exp(-\chi/2)$.
By numerical integration of Eq.
(\ref{eq:vp}) one can  calculate   the electric conductivities 
at finite, non-vanishing 
temperature, for the fields theories dual to CDBBs. 
In particular, we have derived 
the  dependence of 
$\sigma$ on the frequency $\omega$ and on the temperature $T$.
In  the following, we will present the numerical results for $\sigma$
separately 
for the three class of models under consideration.

\subsubsection{Exponential coupling function}

\begin{figure}[ht]
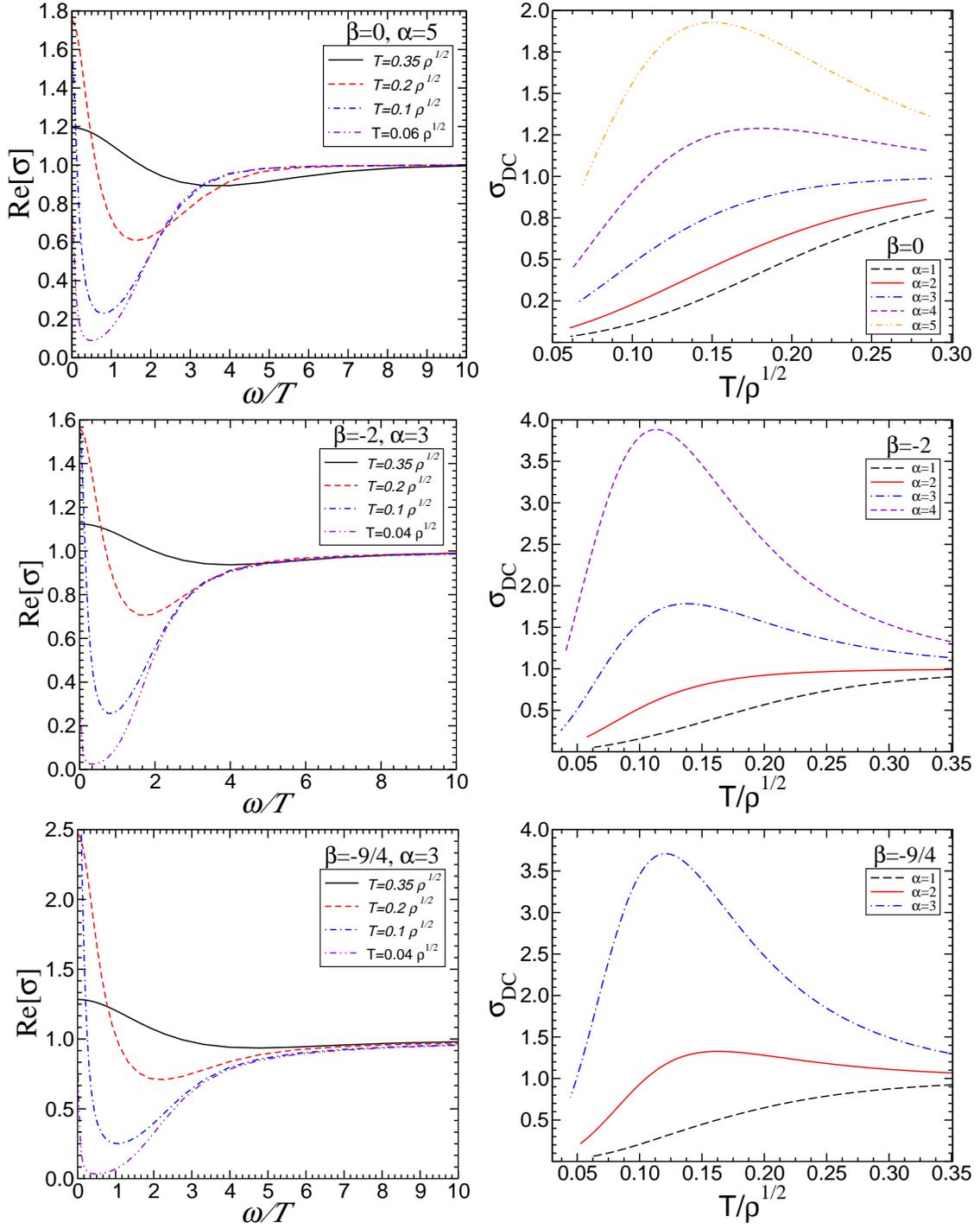

\begin{center}
\begin{tabular}{cc}
\epsfig{file=Plots/CDBB_sigma_VS_omega_beta0.eps,height=6.2cm,angle=0}&
\epsfig{file=Plots/CDBB_sigma_VS_T_beta0.eps,height=6.2cm,angle=0}\\
\epsfig{file=Plots/CDBB_sigma_VS_omega_betam2.eps,height=6.2cm,angle=0}&
\epsfig{file=Plots/CDBB_sigma_VS_T_betam2.eps,height=6.2cm,angle=0}\\
\epsfig{file=Plots/CDBB_sigma_VS_omega_betaBF.eps,height=6.2cm,angle=0}&
\epsfig{file=Plots/CDBB_sigma_VS_T_betaBF.eps,height=6.2cm,angle=0}
\end{tabular}
\caption{Left panels: real part of the AC conductivity as a function
of the frequency for different temperatures for models with 
exponential coupling
function $f(\phi)$ and potential $V(\phi)$ given  by Eqs. 
(\ref{f51}). Right panels: DC
conductivity  as a function of the temperature for several
values of
$\alpha$. Top, middle and bottom panels refer to $\beta=0$,
$\beta=-2$ and $\beta=\beta_{BF}=-9/4$ respectively.
\label{fig:CDBBs_conductivity}}
\end{center}
\end{figure}

Numerical
integration of the equation (\ref{eq:vp}) and the ensuing
calculation of the
conductivity 
(\ref{definition_sigma}) proceeds straightforwardly. A
summary of our results is presented in
Fig.~\ref{fig:CDBBs_conductivity}, where we show the real part of the 
AC conductivity
 for a model with coupling functions $f(\phi)$ and $V(\phi)$ given
by Eq. 
 (\ref{f51})
 as a function of the frequency and the DC conductivity\footnote{As a
consequence of the translation invariance, the imaginary part of
$\sigma$ has a simple pole at $\omega=0$~\cite{Herzog:2009xv}, which
leads to Re$[\sigma]\sim\delta(\omega)$ at $\omega\sim0$. The DC
conductivity is computed by subtracting the Dirac delta
contribution.} as a function of the temperature for selected values
of $\alpha$ and $\beta$.

%
Depending on the values of $\alpha$ and $\beta$, the model
considered shows two interesting features: a ``Drude peak'' in the AC
conductivity at $\omega=0$, a minimum at low frequencies and a
non-monotonic behavior of the DC
conductivity as a function of the temperature. These effects are
absent in the zero temperature limit, where the behavior is mostly
universal. However, they have been previously  observed in models with
$f\sim \cosh(\alpha \phi)$ at $T<T_c$~\cite{Cadoni:2009xm}.
Interestingly, also models with a simple exponential coupling
$f(\phi)$,
which
do not undergo a phase transition AdS-RN/CDBBs and which can be more
easily 
embedded in supergravity theories, show the same
peculiar behavior. 

The appearance of the Drude peak in the AC conductivity 
can be explained in terms of the features  of the 
potential  $V_{s}¥(z)$ entering in the Schr\"{o}dinger equation  that 
determines the reflection coefficient ${\cal R}$ in Eq.~\eqref{sigma}~
\cite{Cadoni:2009xm}. The potential $V_{s}¥(z)$ 
develops a
negative minimum at finite temperature and for large values of
$\alpha$, regardless of the actual form of
$f(\phi)$. As shown in Fig.~\ref{fig:CDBBs_conductivity}, the
critical value of $\alpha$, above which such behavior is manifest,
depends on $\beta$. For $\beta=0$ the maximum of conductivity appears
at $\alpha\gtrsim4$ but, as $\beta$ approaches the BF bound, the
critical value of $\alpha$ is smaller.

Although the conductivity for models with $f\propto \exp(\alpha 
\phi)$ and for models with $f\propto \cosh(\alpha 
\phi)$  behaves qualitatively in the same way,
there are  nevertheless some important differences. 
For the models investigated in Ref.~\cite{Cadoni:2009xm}
the AC conductivity shows a minimum which, for given $\alpha$ and 
$\beta$, is roughly
independent from the temperature.
Conversely, the location of the minimum in
Fig.~\ref{fig:CDBBs_conductivity} depends on the temperature.
Furthermore, at larger frequency the conductivity for the models of
Ref.~\cite{Cadoni:2009xm} develops a maximum before approaching the
universal value $\sigma(\omega\to\infty)\to1$ (cf. Fig.~9 in
Ref.~\cite{Cadoni:2009xm}). Intriguingly, similar transport
properties are observed in the electrical conductivity in graphene
and they are not completely understood in terms of standard theory
(cf. for example Fig.~17 in Ref.~\cite{Peres:2010mx}). On the other
hand, the models with exponential coupling explored in this work do
not show this peculiar behavior.

At relatively small values of $T$ (as those considered in
Fig.~\ref{fig:CDBBs_conductivity}) the resistivity as a function of
the temperature is very well-fitted by
\be
\rho(T)\equiv\frac{1}{\sigma_\text{DC}(T)}=a_0+a_1 T^2+a_2
\log(T)+a_3 T^{-\gamma}\,,\label{fit}
\ee
where $a_i$ are fit parameters depending on $\alpha$ and $\beta$. In
Fig.~\ref{fig:CDBBs_conductivity} these fits are graphically
indistinguishable from numerical data.
The last term in the equation above leads to the following behavior
at small temperatures:
\be
\sigma_\text{DC}(T\sim0)\sim T^\gamma\,.
\ee
When $V(\phi)=-\frac{6}{L^2}\cosh b\phi$, the positive constant
$\gamma$ can be computed by applying the same method discussed in
Sect.~3 of Ref.~\cite{Goldstein:2010aw}, extended to include a
potential as discussed in Ref.~\cite{Cadoni:2009xm}. Its explicit
value reads 
\be
\gamma=2+\frac{b\xi}{1-b\xi}\,,\qquad
\xi=\frac{4(\alpha+b)}{4+(\alpha+b)^2}\,.\label{gammafit}
\ee
Notice that $\gamma=2$ for $\beta=0$~\cite{Goldstein:2010aw} and that
the existence of a near-extremal solution requires
$b\xi<1$~\cite{Cadoni:2009xm}, so that $\gamma>1$. This avoids
reproducing results for strange
metals~\cite{Hartnoll:2009ns,Lee:2010ii,Sachdev:2010uj}, which
typically have
$\gamma=-1$. 
In principle, when different forms of $V(\phi)$ are
considered, the value of $\gamma$ can be computed analytically from
the near-extremal solutions by applying similar arguments (cf.
Sect.~\ref{sect:zerot}).

The fit~\eqref{fit} has been partially inspired by the typical
behavior of the resistivity in ordinary metals. However, a
microscopic description of transport properties in the dual theory is
missing and the fit is only qualitative. In realistic metals, the
first two terms are due to the usual residual resistivity and to the
electron-electron scattering, respectively. The third term is the
celebrated Kondo term, which encloses the non-monotonic behavior at
finite temperature. This term is due to strong coupling interactions
between conduction electrons and impurities. Although no impurities
are present in our model, it is nevertheless intriguing that a
(homogeneous) scalar condensate leads to a non-monotonic behavior of
the resistivity. 

Finally, the last term in Eq.~\eqref{fit} is needed
to correctly reproduce the zero temperature limit and it gives a
vanishing DC conductivity at $T\sim0$. Thus, the non-monotonic
behavior appears to be related to an interplay between the zero
temperature limit (in which $\sigma_\text{DC}$ decreases as $T$ is
lowered) and the finite temperature regime (where, for sufficiently
large values of $\alpha$,  $\sigma_\text{DC}$ increases as $T$ is
lowered). 

Interestingly, the emerging holographic picture
interpolates between some aspects reminiscent of electron motion in
real metals at finite temperature and a charged
plasma~\cite{Goldstein:2010aw} at zero temperature. It would be
highly desirable to develop approximate methods (similar in spirit
to those considered in Ref.~\cite{Goldstein:2010aw}) in order to
extract the exact dependence of the conductivity at finite
temperature and to understand the microscopic mechanism leading to
the observed behavior. We hope to address this issue in a future work.

\subsubsection{Linear coupling function}

\begin{figure}[ht]
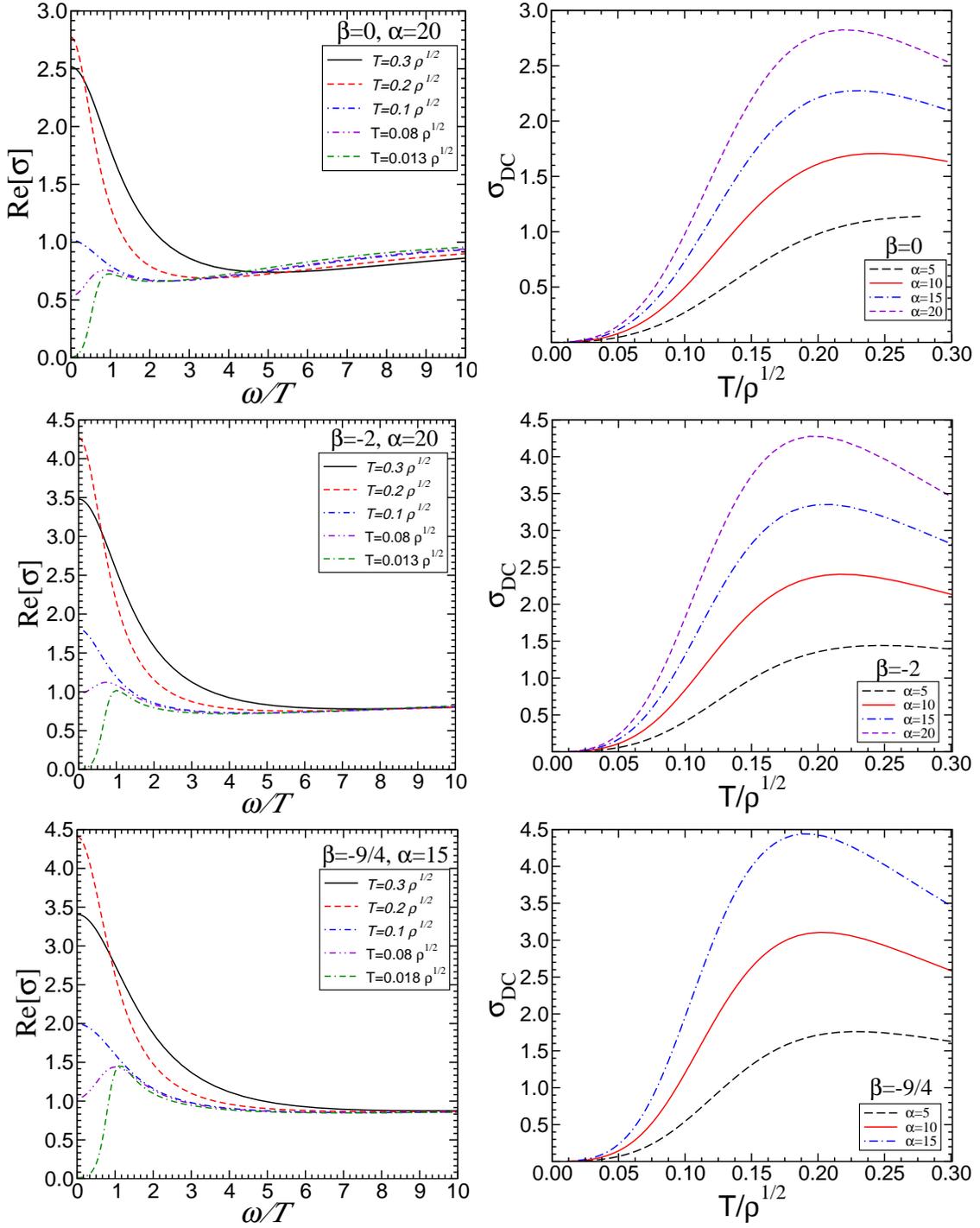

\begin{center}
\begin{tabular}{cc}
\epsfig{file=Plots/CDBB_sigma_VS_omega_beta0_lin.eps,height=6.2cm,angle=0}&
\epsfig{file=Plots/CDBB_sigma_VS_T_beta0_lin.eps,height=6.2cm,angle=0}\\
\epsfig{file=Plots/CDBB_sigma_VS_omega_betam2_lin.eps,height=6.2cm,angle=0}&
\epsfig{file=Plots/CDBB_sigma_VS_T_betam2_lin.eps,height=6.2cm,angle=0}\\
\epsfig{file=Plots/CDBB_sigma_VS_omega_betaBF_lin.eps,height=6.2cm,angle=0}&
\epsfig{file=Plots/CDBB_sigma_VS_T_betaBF_lin.eps,height=6.2cm,angle=0}
\end{tabular}
\caption{Left panels: real part of the AC conductivity as a function
of the frequency for different temperatures for  models with linear
coupling
function $f(\phi)$. 
$f(\phi)$ and $V(\phi)$  are given  by Eq. 
(\ref{f6}) . Right panels: DC
conductivity  as a function of the temperature for several
values of
$\alpha$. Top, middle and bottom panels refer to $\beta=0$,
$\beta=-2$ and $\beta=\beta_{BF}=-9/4$ respectively.
\label{fig:CDBBs_conductivity_lin}}
\end{center}
\end{figure}

Also in the case of linear coupling functions $f(\phi)$ and $V(\phi)$
given in 
Eq.~(\ref{f6}),  the  conductivity of the dual field theory can 
be extracted, numerically, using the method   explained
above. The results are summarized in
Fig.~\ref{fig:CDBBs_conductivity_lin} and are qualitatively similar
to those shown in Fig.~\ref{fig:CDBBs_conductivity}. In particular
the DC conductivity is a non-monotonic function of the temperature
for sufficiently large values of $\alpha$. Also the AC behavior is
similar, but in this case the sharp minimum is smoothed out.
The DC conductivity is again perfectly fitted by Eq.~\eqref{fit}, but
in this case the dominant term in the zero temperature limit is not
$T^\gamma$. We shall compute its exact form in Sect.~\ref{sect:zerot}
(cf. Eq.~\eqref{k1}). 

\subsubsection{Power-law coupling function}\label{power-law1}

\begin{center}
\begin{figure*}[ht]
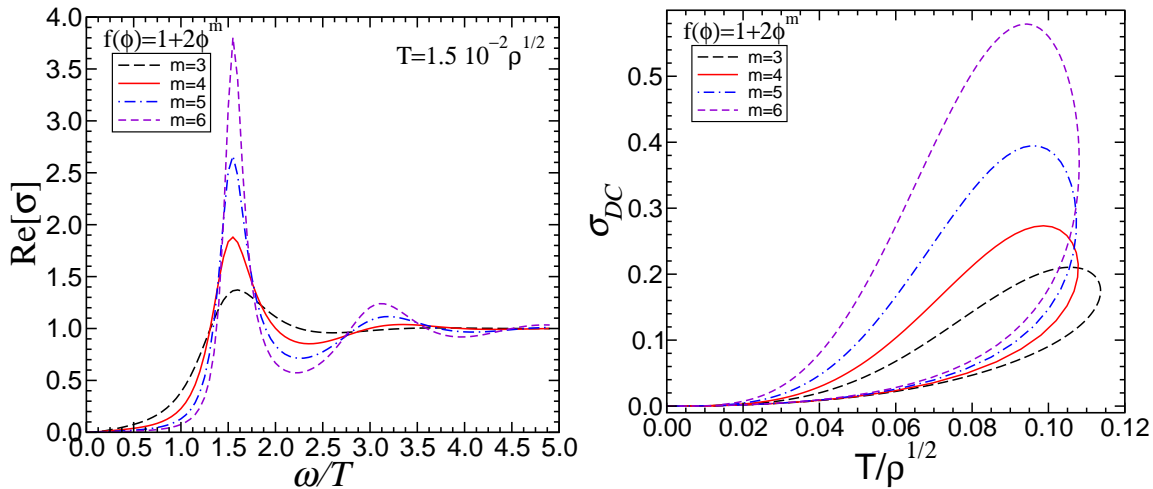

\begin{center}
\begin{tabular}{cc}
\includegraphics[scale=0.32,clip=true]{./Plots/PL_sigmaAC.eps}&
\includegraphics[scale=0.31,clip=true]{./Plots/PL_sigmaDC.eps}
\end{tabular}
\end{center}
\caption{Left panel: Real part of the conductivity as a function of
the frequency for models with  
power-law
coupling function $f(\phi)$ and potential $V(\phi)$ are given by Eq. 
(\ref{f_powerlaw}). We show $Re[\sigma]$  for different values of $m$
and for
$T\sim0.015\sqrt{\rho}$ in the stable branch. Right panel: DC
conductivity as a function of the temperature. Notice that we show
the conductivity for solutions both in the stable and in the unstable
branch.
}
\label{fig:PL_cond1}
\end{figure*}
\end{center}

\begin{center}
\begin{figure*}[ht]
\begin{center}
\begin{tabular}{c}
\includegraphics[scale=0.31,clip=true]{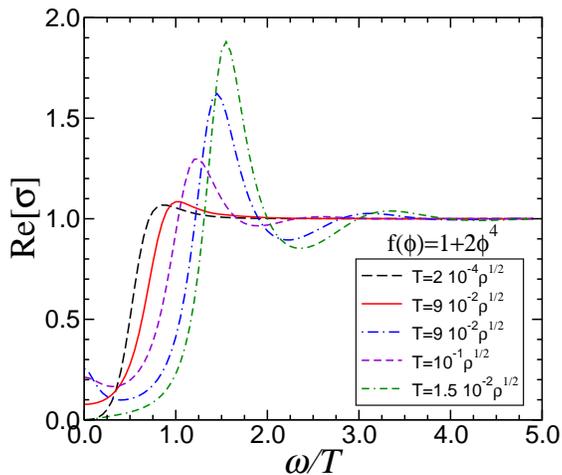}
\end{tabular}
\end{center}
\caption{Real part of the conductivity as a function of the frequency
for $m=4$ and different temperatures for models with power-law 
coupling function $f(\phi)$ and potential $V(\phi)$ are given by
Eq.~(\ref{f_powerlaw}) . The selected
temperatures are
marked in the left panel of Fig.~\ref{fig:PL_O_F} by dots. In
particular notice that we show results both for the stable (dashed
blue) and unstable branch (straight red) for
$T=9\times10^{-2}\sqrt{\rho}$.
}
\label{fig:PL_cond2}
\end{figure*}
\end{center}

The numerical results for the electrical conductivity $\sigma$ in the
dual field 
theory in the case of models with  
coupling functions $f(\phi)$ and $V(\phi)$ given by
Eq.~(\ref{f_powerlaw}) are 
shown in 
Figs.~\ref{fig:PL_cond1} and \ref{fig:PL_cond2}.

In the left panel of Fig.~\ref{fig:PL_cond1} we show the AC
conductivity
at $T\sim0.015\sqrt{\rho}$ for different values of $m$ and for
solutions in the stable branch. Sharp peaks
develop at the same frequency $\omega_\text{peak}\sim1.5T$, which is
largely independent from $m$, while the height of the peak increases
with $m$. In the right panel of Fig.~\ref{fig:PL_cond1} we shown the
DC conductivity as a function of the temperature for solutions both
in the stable and in the unstable branch, for different values
of $m$. Again the qualitative behavior is very similar. Notice that,
a part from the peculiar behavior of the temperature, also in this
case the conductivity has a non-monotonic behavior. As we prove in
Sect.~\ref{sect:zerot}, in the zero temperature limit the DC
conductivity universally approaches zero with a power-law behavior.

Finally, in Fig.~\ref{fig:PL_cond2}, we show the behavior of 
the AC conductivity  as a a function of $\omega$ for  $m=4$ and for 
selected values of the temperature.

\section{The Zero temperature limit of electric charged black branes}
\label{sect:zerot}
In the previous section we have constructed numerical solutions  
for  several CDBBs  at finite, nonvanishing, temperature and 
investigated some of their holographic features.
In this section we will discuss  the zero temperature behavior 
of  our models, by considering the near-horizon regime of the 
black brane  solutions.
The near-horizon, extremal 
behavior of charged dilaton black holes for which  the
 coupling 
function and/or  the  potential behave exponentially,   has been
already investigated in Refs.
\cite{Goldstein:2009cv,Goldstein:2010aw,Cadoni:2009xm,Perlmutter:2010qu}.
In this 
section we will extend this analysis to include the case of a
power-law 
behavior of $f(\phi)$ and/or $V(\phi)$.

A general problem one has to tackle while searching for extremal 
near-horizon solutions of AdS black branes is the possibility of 
connecting smoothly these solutions with the asymptotic AdS$_{4}$ 
region. In other words, the near-horizon solutions should 
allow for  
the existence of a global black brane solution interpolating between  
them and 
the asymptotic AdS$_{4}$ region. 
It is not easy to perform this analysis analytically. 
A possibility is to use the method proposed in Ref. 
\cite{Goldstein:2009cv}  (see also Ref. \cite{Cadoni:2009xm}).
A subleading deformation
term, which  
grows  outside the near-horizon region and becomes dominant in the 
asymptotic region giving the correct AdS$_{4}$ behavior, 
is introduced in  the near-horizon solution .

For generic EMDG models this method is 
not easy to implement, mainly because the subleading terms in 
the near-horizon expansion of the metric and scalar field are
difficult to control.

For this reason in this paper we will use a simplified approach: we 
will just compute, using the field equations, the leading term in the 
near-horizon expansion of the  bulk fields. These leading term will 
be used to investigate the holographic features of the dual field 
theory. The question about the existence of global solutions 
interpolating between the leading near-horizon solution  and the
asymptotic 
AdS$_{4}$  behavior will be tackled numerically for a number of
cases.   

\subsection{The near-horizon, extremal solution}

We can assume without loss of generality that the horizon of our 
extremal CDBB solutions is located 
at $r=0$. We are  interested in the leading term of the asymptotic 
expansion of the solution near $r=0$. Because in the near-horizon 
region we expect the theory to be strong-coupled we  assume that 
$\phi\to\infty$ as $r\to 0$. Hence  the extremal CDBB solutions are 
determined by   the $\phi\to\infty$ asymptotic form of the functions
$f(\phi)$ 
and $V(\phi)$.

To keep the discussion as general as possible, we  will  consider 
coupling functions with the following leading term in the 
$\phi\to\infty$ expansion: 

\ba\label{f1}
{\rm a})\quad V(\phi)&=a\phi^{n}&,\quad f(\phi)=b\phi^{m};\nonumber\\
{\rm b}) \quad V(\phi)&=a\phi^{n}&,
\quad f(\phi)=be^{\alpha \phi};\nonumber\\
{\rm c}) \quad V(\phi)&=a e^{\alpha \phi}&,
\quad f(\phi)=b \phi^{m};
\ea
Where $a,b,n\ge 0,m\ge 0$ and $\alpha\ge 0$ are constants.

Notice that the  leading term in the $\phi\to\infty$ expansion of the
exponential, linear, power-law coupling functions $f(\phi)$ 
and the self-interaction potentials $V(\phi)$ considered in 
Sect.~\ref{sec:CDBBs} are all particular cases of Eqs. 
(\ref{f1}).

In this section we use the following parametrization for the
metric 
\be
ds^2 = - \lambda(r) dt^2 + \frac{dr^2}{\lambda(r)} +
H^2(r) (dx^2 + dy^2)\label{ansatz_T0} \ . \ee
The equation of motion for the gauge
field can be immediately integrated and gives \be A_0' = \frac{\r}{f
H^2} \ , \label{cond:gaugefield}
\ee where $\r$ is the charge density
of the solution.
The remaining equations are 
\ba\label{e2} 
(\l H^2)'' &=& -2   H^2 V(\phi) \, \nb \\
(H)'' &=& - \frac{H}{4}  (\phi')^2\, \nb \\
(\l H^2 \phi')' &=&  H^2\frac{dV}{d\phi} -
\frac{\r^2}{2H^{2}}\frac{1}{f^{2}}\frac{d f}{d\phi}  \, \nb \\
\l(H')^2 + \frac{\l'}{2}(H^2)' &=&  \frac{H^2}{4} \left [\l
(\phi')^2 -  \frac{\r^{2}}{ f H^{4}}  - 2 V  \right ] 
. 
\ea 
To find the leading term in the 
near-horizon  expansion of the fields, we  try the
following  
scaling ansatz 
\be\label{scalingan}
\phi=\phi_{0}(-\ln r)^{h},\qquad 
H=C r^{\nu}(-\ln r)^{p},\qquad \l=\l_{0} r^{\mu}(-\ln r)^{q},
\ee
where $\phi_{0},h,C,\nu,p,\l_{0},\mu,q$ are constants. 
We discuss separately the three cases under consideration.

\subsubsection{Power-law coupling functions $f(\phi)$ and $V(\phi)$} 
\label{a)}

Using the scaling ansatz (\ref{scalingan}) in the  field equations
(\ref{e2}) 
one finds  in this case the leading behavior of the near-horizon 
solution:
\be\label{e3}
H=C (-\ln r)^{-\frac{n+m}{8}},\quad \l= - a\phi_{0}^{n}r^{2}(-\ln 
r)^{\frac{n}{2}},\quad \phi=\phi_{0}(-\ln r)^{\frac{1}{2}},
\ee
where 
\be\label{e4}
\phi_{0}= \left[2\left(n+m\right)\right]^{1/2},\quad\quad
C^{4}=-\frac{\r^{2}}{2 
ab}\left[2\left(n+m\right)\right]^{-\frac{n+m}{2}}.
\ee
Notice that the parameter $a$ must be negative and $b$ positive.
For $n=0$ this is consistent 
with the presence of a negative cosmological constant in the action, 
$a=-\frac{6}{L^{2}}$. In the case of a quadratic $V(\phi)$,
$a=m_{s}^{2}/2$, so 
that the scalar squared mass must be negative.

Using the field equations one can easily compute the leading 
behavior of the scalar curvature tensors on the horizon:
\be\label{e5}
R= 2 a\phi_{0}^{n} (-\ln r)^{\frac{n}{2}},\quad R_{\mu\nu} R^{\mu 
\nu}\sim  (-\ln r)^n.
\ee

The leading term of the scalar curvatures does not depend on the 
coupling function $f(\phi)$, i.e. it depends on $n$ but not on $m$.
For $n=0$, independently  of $m$, the scalar curvature  $R$ on the
horizon 
goes to a constant,  which is essentially the cosmological constant
of 
the spacetime. For $n\neq 0$, we have in $r=0$ a mild logarithmic  
singularity $R\sim (- \ln r)^{n/2}$.

\subsubsection{ Exponential $f(\phi)$ and power-law $V(\phi)$}
\label{b)}

The field equations (\ref{e2}) are now solved at leading order
by Eqs.~(\ref{e3}), (\ref{e4})  with $m=0$. The only difference is
that now the 
term depending on $f(\phi)$ in the third equation (\ref{e2}) is
subleading 
with respect to the term depending on $V(\phi)$, so that $C$ is not
anymore 
determined by Eq.~(\ref{e4}) but is a free integration constant. 

\subsubsection{ Power-law $f(\phi)$ and exponential $V(\phi)$}
\label{ c)}

In this case the field equations (\ref{e2}) allow for  metric 
solutions  having a near-horizon Lifshitz scaling  behavior and 
logarithmic scalar field.   
These Lifshitz-like solutions exist when $0< \alpha <2$ . They read

\be\label{e7}
H=C r^{\gamma},\quad \l= -\frac{\alpha^{2} 
a}{4\gamma(1-2\gamma)}r^{2-4\gamma},\quad 
\phi=-\frac{4\gamma}{\alpha}\ln r,\quad 
\gamma=\frac{\alpha^{2}}{4+\alpha^{2}},
\ee
where  $C$ is a free integration constant.
It is important to stress that the solution does not depend on the 
coupling function $f(\phi)$, i.e. on the parameters $b,m$. This is
because 
the  term depending on $f(\phi)$ in the third equation (\ref{e2}) is
either
subleading 
with respect to the term depending on $V(\phi)$ ($m\neq 0$) or
identically 
zero ($m= 0$).
In particular, for $m=0$ ($f=b$) our model describes a scalar field  
minimally coupled  with the $U(1)$ field, with exponential potential.
This is an important result. It shows that extremal solutions with
Lifshitz
scaling are not only solutions of  dilaton gravity models with 
exponential coupling function $f(\phi)$ 
\cite{Goldstein:2009cv,Goldstein:2010aw,Cadoni:2009xm}, but can be
also 
obtained in the case of a minimally coupled dilaton 
\footnote{Actually  our 
solution  (\ref{e7}) can be obtained also as a particular case of the 
solution of \cite{Cadoni:2009xm},  by setting the coupling function 
$f(\phi)$ to a constant.}.

\subsubsection{Exact solutions at $T=0$}
In some case Eqs.~(\ref{e2}) allow also for exact $AdS_2\times R^2$
solutions. Let us that assume $V(\phi)=V_0+a \phi^n$ and
$f(\phi)=f_0+b \phi^m$, where $n,m>0$ and, in order to recover an
asymptotic $AdS_4$, we impose $V_0=-6/L^2$ and $f_0=1$. Then, there
exists an exact solution which reads
\be
\lambda(r)=\lambda_0r^2=-(V_0+a\phi_0^n) r^2\,,\qquad
H(r)=H_0=\left(\frac{\rho^2 b
m\phi_0^{m-n}}{2an(f_0+b\phi_0^m)^2}\right)^{1/4}\,,\qquad
\phi(r)=\phi_0\,,\label{solT0}
\ee
provided $\phi_0$ is an extremum of the effective potential 
$V_{eff}=H_{0}^{2}V+(\r/2H_{0}^{2})f^{-1}$, i.e. it is solution of
the following equation
\be
ab(m+n)\phi_0^n+a f_0n\phi_0^{n-m}+bm V_0=0\,.\label{eqphi0}
\ee
Notice that Eqs.~\eqref{solT0} imply $a,b<0$. For example if $n=m=2$,
then $a=m^2_s/2<0$ and $b=\alpha<0$ has the opposite sign with respect
to the case studied in Ref.~\cite{Cadoni:2009xm}.

In some particular case the equation above can be solved
analytically. For example if $n=m$ we have
\be
\phi_0^n=-\frac{bV_0+af_0}{2ab}\,.\label{eqphi02}
\ee
Let us close this subsection by stressing the fact that the solutions 
we have presented here not only give the leading term of  $T=0$ 
extremal black branes, but can be generically  used also  in the 
near-extremal region 
$T\sim 0$. Thus, our Eqs.~\eqref{e3} and \eqref{e7} can be used to
construct also
the 
 the leading behavior of near-extremal  black branes, as explained
e.g. in Ref.~\cite{Goldstein:2009cv}. In this case 
 the horizon radius $r_{h}$ is small but  nonvanishing. 

\subsection{Holographic properties}
In this section we will discuss the transport features of the field 
theory dual to extremal CDBBs at $T\sim 0$, using the the
Schr\"odinger-like 
picture for the dynamics of electric bulk perturbations
 described at the beginning of Sect.~\ref{holoprop} 
and, in particular, Eq.~(\ref{sigma}).

The Schr\"odinger-like description is particularly useful when one 
has bulk solutions written in explicit analytic, albeit approximate, 
form. This is because from the near-horizon behavior  of the 
solution one can derive  the near-horizon behavior of the 
potential $V_{s}¥(z)$ of the Schr\"odinger equation.
Assuming that the
near-horizon solution can be smoothly connected with the asymptotic
$AdS_4$,
the leading power in $\omega$ of the conductivity $\sigma$ 
(\ref{sigma}) can be 
calculated by matching the conserved probability current of the 
Schr\"odinger
equation near the boundary at infinity and near the
horizon~\cite{Goldstein:2009cv}.

The Schr\"odinger-like equation  and the  related  potential 
read \cite{Cadoni:2009xm},
\be\label{schrod}
\frac{d^2\Psi}{dz^2}+(\omega^2-V_{s}¥(z))\Psi=0,\quad 
\Psi=\sqrt{f}A_{x},\quad
\frac{dr}{dz}=\lambda,\quad
V_{s}¥(z)=\lambda
f(\phi){(A_0')}^2+\frac{1}{\sqrt{f(\phi)}}\frac{d^2\sqrt{f(\phi)}}{dz^2}\,.
\ee
The case of the Lifshitz-like solutions (\ref{e7}) has been already 
investigated in
Ref.~\cite{Goldstein:2009cv,Cadoni:2009xm,Goldstein:2010aw} (see also
Refs.~\cite{Bertoldi:2010ca,Bertoldi:2011zr}).
A near-horizon  $V_{s}¥(z)\sim {C}/{z^{2}}$ was found, which in 
general gives a leading term for $\sigma$ scaling as some power of 
$\omega$, namely $\sigma \sim \omega^{s}$ with $s\ge 2$
\cite{Cadoni:2009xm}. 

Let us now consider the solution (\ref{e3}) corresponding to  the
near-horizon behavior 
of  CDBBs with power-law coupling $f(\phi)$. Notice that this 
solution contains as particular case, $m=1$, a linear coupling 
function $f(\phi)$.
The Schr\"odinger potential $V_{s}¥(z)$ is easy calculated using
Eqs.~(\ref{e3}) into Eq.~(\ref{schrod}). 
It reads
\be
V_{s}¥(z)\sim\frac{2}{z^2}\,,
\ee
independently from the values of $m$ and $n$. Hence, assuming the
near-horizon solution can be smoothly connected with the asymptotic
$AdS_4$, the same analysis of
Refs.~\cite{Goldstein:2009cv,Cadoni:2009xm,Goldstein:2010aw} applies
and the conductivity reads
\be
\text{Re}[\sigma(T\sim0)]\sim\omega^2\,, \label{cond_T0}
\ee
for any $m$ and $n$ and when $\omega\sim0$.
This is an important result: in the $T\sim 0$ regime 
the optical conductivity as a function  of the frequency  for the
field 
theories holographically dual to CDBBs with power-law coupling 
functions has a universal quadratic  scaling behavior. 

Let us now consider  the DC conductivity. At $T\sim0$ the DC 
conductivity as a function of $T$  can be computed by
applying the same procedure used  in
Ref.~\cite{Goldstein:2010aw}. In the near-horizon 
region $\text{Re}[\sigma]$ is first expressed as a function of the
horizon
radius 
$r_{h}$, then as function of $T$ by means of  the radius-temperature 
relation, $r_{h}(T)$. 
In the $T\to0$ limit, the near horizon behavior of our numerical
solutions is consistent with $\lambda\sim\lambda_0 r^2[-\log
r]^{n/2}\left[1-\left({r}/{r_h}\right)^\eta\right]$, where $\eta$
depends on the details of the model. Hence, at
leading order, the radius-temperature 
relation reads
\be
T\sim r_h\left[-\log r_h\right]^{{n}/{2}}\,,\label{TVSrh}
\ee
which can be formally inverted to obtain $r_h(T)$. This implies 
\be\label{k1}
\sigma_\text{DC}\equiv\text{Re}[\sigma(\omega\sim0)]\sim r_h^2\sim
{T^2}\left\{-\log\left[r_h(T)\right]\right\}^{-{n}/{2}}\,,
\ee
where the first relation follows from the arguments in
Ref.~\cite{Goldstein:2010aw}, whereas the second one follows from
Eq.~\eqref{TVSrh}.

The DC conductivity does not depend on the nonminimal coupling
$f(\phi)$, but only on the self-interaction potential $V(\phi)$. When
$n=0$ we
recover the ``universal'' behavior $\sigma_\text{DC}\sim T^2$, while
for $n\neq0$, although the relation~\eqref{TVSrh} cannot be inverted
analytically, we expect some subdominant contribution. 

Let us now consider the case in which the near-horizon solution is 
$AdS_2\times R^2$ and is described by Eq.~(\ref{solT0})
Interestingly, it follows from
Eqs.~\eqref{solT0}, \eqref{eqphi0} and \eqref{schrod}
that also in
this case the Schr\"odinger potential near the horizon reads
$V_{s}¥(z)\sim2/z^2$, regardless of the values of $m$ and $n$. 
Moreover in this case the
temperature scales linearly with the radius, $T\sim r_h$. Thus, as
explained in Refs.~\cite{Goldstein:2009cv,Cadoni:2009xm}, we find the
same ``universal'' behavior found in Ref.~\cite{Goldstein:2010aw}
\be
\text{Re}[\sigma(T\sim0)]\sim\omega^2\,, \qquad
\sigma_\text{DC}\equiv\text{Re}[\sigma(\omega\sim0)]\sim T^2\,.
\ee
\subsubsection{Comparison with the numerical results}
Let us now compare the analytical results for the $T\sim 0$ region
described in this section with the numerical results 
of Sect.~\ref{sec:CDBBs}.

The first issue to be discussed concerns the possibility of jointing 
smoothly the near-horizon, extremal, approximate solutions
(\ref{e3})-(\ref{e7}) 
with the asymptotic AdS$_{4}$  form of  the solutions. We have not
performed such analysis analytically, but we have done it numerically
by a case by case analysis. Indeed, for the various cases presented
in this section, we have explicitly 
checked that the near-horizon limit of the numerical CDBBs discussed
in Sect.~\ref{sec:CDBBs} approaches the analytical
behavior~\eqref{scalingan} at leading order in the $T\to0$ limit.

A second important issue is the comparison of our numerical results 
for the AC conductivity $\sigma_{AC}(\omega)$ and for the DC 
conductivity $\sigma_{DC}(T)$ obtained   in Sect.~\ref{sec:CDBBs}
with,  respectively 
Eqs.~(\ref{cond_T0}) and (\ref{k1}). In the $T\to0$
limit, we have explicitly checked for the case of 
a linear and power-law coupling function that our numerical results
for $\sigma_{AC}(\omega)$  and  
$\sigma_{DC}(T)$
presented in Sect.~\ref{sec:CDBBs} agree respectively with 
Eqs.~\eqref{cond_T0} and (\ref{k1}). As a general results, our
numerical solutions and the analytic expectations at $T\sim0$ are in
full agreement, validating each other.
\section{Dyonic black branes at finite temperature}\label{sec:DDBBs}
In this section, we shall consider solutions of the
theory~\eqref{lagr_genB}, which describe \emph{dyonic} dilatonic
black branes (DDBBs), i.e. dilatonic solutions endowed with both an
electrical and a magnetic charge. 

The zero temperature limit of DDBBs
have been studied in detail in Ref.~\cite{Goldstein:2010aw} in the
case of exponential coupling and for a SL(2,R) invariant action
including an axion field. In that case dyonic solutions can be
constructed from purely electrical solutions by applying the
electromagnetic duality~\cite{Goldstein:2010aw}.  We
shall restrict ourself to the theory~\eqref{lagr_genB}, which is not
SL(2,R) invariant. Moreover, we shall consider couplings such
that $f'(0)=0$, i.e. theories allowing for dyonic AdS-RN black
branes. Our main motivation is to understand the role played by the
magnetic
field in the phase transition and in the holographic properties 
of the dilatonic black branes investigated  in
Ref.~\cite{Cadoni:2009xm}.

As we have argued  in the previous sections, the holographic
properties of dilatonic black branes are \emph{qualitatively} similar
regardless of the details of the coupling. Hence, we expect that the
results we discuss below apply to a broader class of
EMDG models. Finally, for completeness in
Appendix~\ref{appendix} we discuss purely magnetic black brane
solutions obtained in our theory via the electromagnetic duality.

In the following, we consider a dyonic configuration for the gauge 
potential
$A=A_\mu dx^\mu= A_0(r)dt+Bx dy$, where $B$ is the magnetic field,
together with the ansatz 
Eq.~\eqref{metric_ansatz_BR} for the metric. The first and fourth of
Eqs.~\eqref{eq:BR_einstein} then read
\beq
\phi''+\left(\frac{g'}{g}-\frac{\chi'}{2}+\frac{2}{r}\right)\phi'(r)-\frac{1}{g}\frac{d
V}{d \phi}+\frac{1}{2g} \frac{d f}{d
\phi}\left({A_0'}^2e^\chi-\frac{B^2}{r^4}\right)&=&0\,,\label{eq:BR_scalarB}\\
\frac{{\phi'}^2}{4}+\frac{f(\phi)}{4g}\left({A_0'}^2e^\chi+\frac{B^2}{r^4}\right)+\frac{g'}{rg}+\frac{1}{r^2}+\frac{V(\phi)}{2g}&=&0\,,
\label{eq:BR_einstein1B} \eeq 
whereas the second and the third of Eqs.~\eqref{eq:BR_einstein} are
not affected by $B$. We look for solutions of these four coupled
nonlinear ODEs which describe a static, planar black brane, endowed
with an electric field, a magnetic field
perpendicular to the $(x,y)$ plane and a scalar field. Notice that
these solutions are translationally invariant 
in the $(x,y)$ direction, unlike those
obtained by considering a minimal
coupling~\cite{Albash:2008eh,Montull:2009fe}. 

Finally, we are interested in models which
admit AdS-RN black branes as solution. For this purpose, we restrict
ourself to the following form for the potential and the nonminimal
coupling:
\be
V(\phi)=-\frac{6}{L^2}+\frac{\b}{2L^2}\phi^2+{\cal O}(\phi^3)\,,\qquad
f(\phi)=1+\frac{\alpha}{2}\phi^2+{\cal O}(\phi^3)\,.
\label{expVandfB} 
\ee
Due to the expansions~\eqref{expVandfB}, the dyonic AdS-RN black
brane is
solution of the equations of motion with
\be
g\equiv
g_{RS}=-\frac{2M}{r}+\frac{Q^2+B^2}{4r^2}+\frac{r^2}{L^2}\,,\qquad
\chi = 0
 \,,\qquad
A=\left(\frac{Q}{r}-\frac{Q}{r_h}\right)dt+Bxdy\,, \qquad \phi = 0
\,. \label{eq:RSAdSB} 
\ee
%
\subsection{Instability of dyonic AdS-RN black branes}
When $B=0$, below a critical temperature the AdS-RN solution is
unstable against scalar
perturbations~\cite{Cadoni:2009xm}.
Following Ref.~\cite{Cadoni:2009xm}, we consider scalar perturbations
around the AdS-RN black
brane~\eqref{eq:RSAdSB}, and Fourier-expanding the perturbation as
$\phi_{\w, \vec{k}} =
\frac{R(r)}{r} e^{i (k_1 x + k_2 y - \omega t)}$, we find a
Schr\"odinger-like equation 
\be g_{RS}^2 R''+g_{RS} g_{RS}'R' +\left[\omega^2-V_{s}(r)
\right]R=0\,, \qquad V_{s}(r) = g_{RS}
\left[\frac{\vec{k}^2}{r^2} + \frac{g_{RS}'}{r} +
m^2_{\text{eff}}\right]\,, \label{scalar_pertB}  \ee
where the effective mass reads
%
\be 
m^2_{\text{eff}}(r) =
m_{s}^2-\alpha\frac{Q^2-B^2}{2r^4}\,,\label{eff_massB}
\ee
and again $m_{s}^2=\beta/L^2$ is the squared mass of the scalar field.
Interestingly, the contributions of the magnetic and of the electric
field are opposite. While the electric field contributes to a
tachyonic mode in the effective mass (hereafter we focus on
$\alpha>0$), the magnetic field gives a \emph{positive} contribution
and stabilizes the AdS-RN black brane. For dyonic black branes, these
two contributions are competitive and an instability can arise only
when $Q^2>B^2$.
Indeed, the effective square mass is positive above a critical value
of
the magnetic field, i.e. for $B>B_c$ where
\be
B_c=\pm Q\,.\label{Bc}
\ee
Independently from the value of $\alpha$, dyonic AdS-RN black branes
with $B\geq B_c=Q$, are stable against scalar perturbations. 
On the other hand, for $B<Q$ 
the non-minimal coupling gives a 
negative contribution to the effective
mass. If the coupling is strong enough
it can lower the mass below the BF bound and destabilize the
background.
In the
following, we shall confirm this result by solving numerically the
equations of motion. 
\subsection{Dyonic dilatonic black branes and phase transitions}
DDBB solutions of
Eqs.~\eqref{eq:BR_scalarB}-\eqref{eq:BR_einstein1B} can be obtained
numerically by using the same procedure sketched in the previous
sections and discussed in detail in
Ref.~\cite{Cadoni:2009xm}. 
When $B=0$, a phase transition occurs below a critical
temperature, the two phases being the (unstable) AdS-RN black brane
and the new (stable) charged dilatonic black brane. Here we want to
understand the role of the magnetic field in the phase
transition\footnote{See also Ref.~\cite{Lugo:2010qq} for phase
transitions from
dyonic solutions in Einstein-Yang-Mills-Higgs theory.}.

The black brane temperature $T$ is simply defined in terms of
near-horizon quantities as $(L=1)$
\be
T=\frac{g_h'}{4\pi}e^{-\frac{\chi_h}{2}}=\frac{r_he^{-\frac{\chi_h}
{2}}}{4\pi}\left(\frac{1}{r_h^2}-e^{\chi_h}\frac{A_0(r_h)^2f(\phi_h)}
{4}-\frac{V(\phi_h)}{2}-\frac{B^2
f(\phi_h)}{4 r_h^2}\right)\,.
\ee
As previously explained, when $B=0$ the solutions form a
one-parameter family, the physical parameter being the temperature
$T$. However when $B\neq0$, the series expansion near the horizon
depends
on three independent parameters, say $\phi_h$,
$A_0(r_h)$ and $B$. At given $\phi_h$, the remaining parameters can
be chosen to be the zeros of two
functions of two variables,
\be
\left\{
\begin{array}{ll}
 \displaystyle 
{\cal F}_1:\,\left\{A_0(r_h),B\right\}\longrightarrow {\cal
O}_{i}\,,\\
{\cal F}_2:\,\left\{A_0(r_h),B\right\}\longrightarrow {B}/{\rho}-C\,,
\end{array}\right.
\label{root1}
\ee
where $\rho$ is the charge density defined by the asymptotic
expansion of the gauge field at infinity. Both
${\cal O}_{i}$ and $B/\rho$ result from the numerical integration and
$C$ is the value of the constant magnetic field in units of $\rho$.
Therefore numerical solutions form a two-parameter family: the
parameters can be chosen to be
the black brane temperature $T/\sqrt{\rho}$ and the magnetic field
$C=B/\rho$. The functions ${\cal F}_i$ are known only numerically and
they are not necessarily polynomials. In order to find their zeros we
implemented a two-dimensional extension of M\"uller method, recently
proposed in Ref.~\cite{Fiziev:2010yy}. Solving the
system~\eqref{root1} iteratively (i.e. for several values of
$\phi_h$), we can follow the evolution of the condensate as a
function of the temperature at fixed magnetic field. Alternatively,
we can also study the evolution of the condensate as a function of
the magnetic field at fixed temperature. This is necessary to study
the magnetic susceptibility, as we explain below. In this case, we
should find the zeros of the following functions
\be
\left\{
\begin{array}{ll}
 \displaystyle 
{\cal F}_1:\,\left\{A_0(r_h),B\right\}\longrightarrow {\cal
O}_{i}\,,\\
{\cal F}_2:\,\left\{A_0(r_h),B\right\}\longrightarrow
{T}/{\sqrt{\rho}}-C\,,
\end{array}\right.
\label{root2}
\ee
where now $C$ is the constant temperature in units of $\sqrt{\rho}$. 
\subsubsection{Numerical results}
We refer to Refs.~\cite{Fiziev:2010yy,Cadoni:2009xm} for further
details on the numerical method, now focusing on some results.
Consistently with the requirements~\eqref{expVandfB}, we have
considered both polynomial forms $V,f\sim a+b\phi^2$ and hyperbolic
cosine forms $V,f\sim \cosh(a\phi)$ for the potential $V(\phi)$ and
the coupling function $f(\phi)$. Results are qualitatively similar
regardless of the precise
form of $f,V$ and they only show a strong dependence on
$\beta \equiv V''(0)L^2$ and $\alpha\equiv
f''(0)$. 

For concreteness, we shall focus on
$\beta=m_{s}^2L^2=-2$ and on solutions obtained by imposing ${\cal
O}_-=0$. 
Imposing ${\cal O}_+=0$ or choosing different values of the scalar
mass
$\beta$, gives qualitatively similar results. 

\begin{center}
\begin{figure*}[ht]
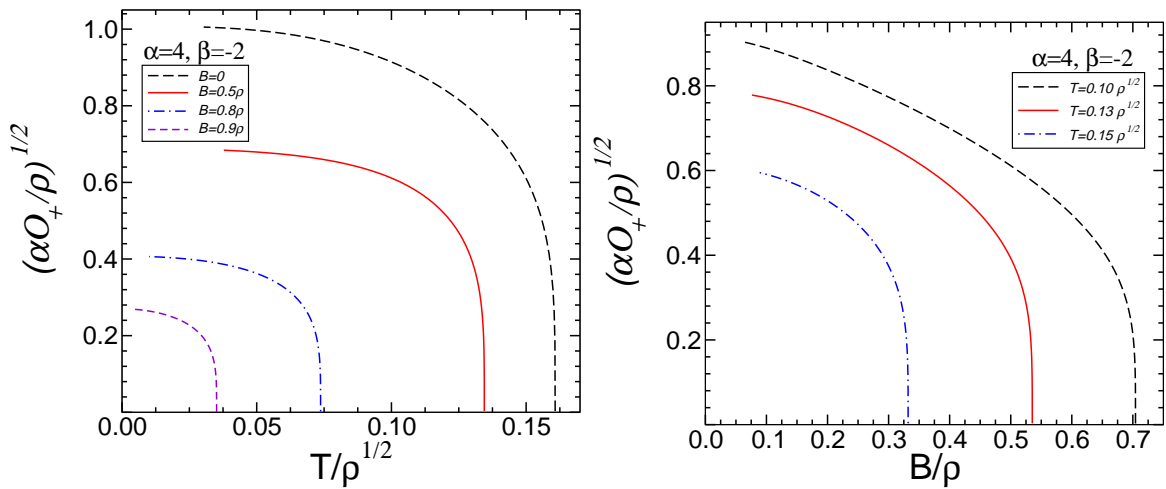

\begin{center}
\begin{tabular}{cc}
\includegraphics[scale=0.31,clip=true]{./Plots/O2_VS_T.eps}&
\includegraphics[scale=0.31,clip=true]{./Plots/O2_VS_B.eps}
\end{tabular}
\end{center}
\caption{Left panel: scalar condensate for DDBBs with as a function
of the
temperature for selected values of the magnetic field $B$. Right
panel: scalar condensate as a function of the magnetic field for
selected values of the temperature. 
We used $f(\phi)=\cosh(2\phi)$ and $\beta=-2$.
}
\label{fig:condensateB}
\end{figure*}
\end{center}

\begin{center}
\begin{figure*}[ht]
\begin{center}
\begin{tabular}{c}
\includegraphics[scale=0.31,clip=true]{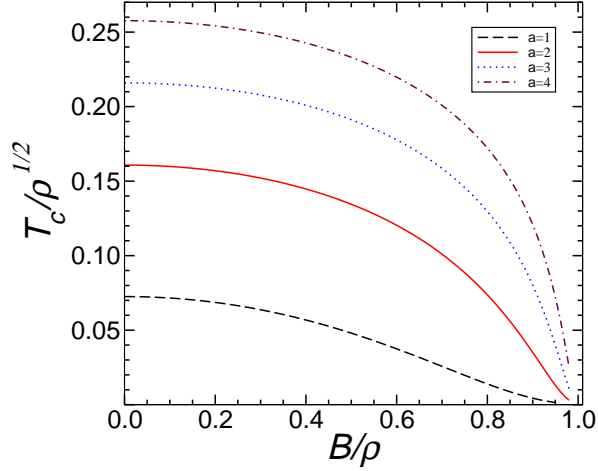}
\end{tabular}
\end{center}
\caption{Phase diagram for dyonic and dilatonic black branes  with
$f(\phi)=\cosh(a\phi)$, showing
the critical temperature as a function of the ratio $B/\rho$. 
Regions below the curves mark the parameter space where the
scalar operator condenses. We
considered  several values of
$a=\sqrt{\alpha}$. Choosing $f=1+{\alpha}\phi^2$ gives qualitatively
similar results.}
\label{fig:Tc_VS_B}
\end{figure*}
\end{center}

First, we report that this method is successful in constructing
dyonic AdS black branes coupled to a neutral scalar field. From the
holographic point
of view, DDBBs are dual to field theories in which a neutral scalar
operator acquires a non-vanishing expectation value below a critical
temperature \emph{and below a critical magnetic field}. In fact, in
Fig.~\ref{fig:condensateB} we show the scalar condensate both as a
function of the temperature for several values of constant magnetic
field (left panel) and as a function of the magnetic field for
several values of constant temperature (right panel) for 
$\alpha=4$ and $\beta=-2$. 
The phase diagram of DDBBs, i.e. the critical
temperature as a function of the magnetic field, is shown in
Fig.~\ref{fig:Tc_VS_B} for  several values of the coupling $\alpha$.
The numerical results confirms our analytic expectation: $T_c=0$ when
$B=B_c=\rho$, i.e. the scalar operator does not condense at any
finite temperature above the critical magnetic field $B_c$. The
critical value $B_c$ does not depend neither on the coupling constant
$\alpha$ nor on the precise forms for $f(\phi)$ and $V(\phi)$, when
they behave as prescribed in Eq.~\eqref{expVandfB}.
Remarkably, Fig.~\ref{fig:Tc_VS_B} is qualitatively similar to Fig.~8
in Ref.~\cite{Hartnoll:2008kx}, which describes the (qualitative)
phase diagram for holographic superconductors immersed in a magnetic
field. 

The phase diagram for DDBBs shown in Fig.~\ref{fig:Tc_VS_B} is exact
and confirms the schematic illustration depicted in
Ref.~\cite{Hartnoll:2008kx}.

Another interesting issue is the nature of the phase transition shown
in the right panel of Fig.~\ref{fig:condensateB}. At $B\sim B_c$ the
scalar condensate has the typical behavior for second order phase
transitions in the mean-field approximation,
\be
O_i\sim(B-B_c)^{1/2}\,. \label{eq:typeII}
\ee
This kind of phase transitions occurs in type-II superconductors. The
same behavior is observed in real high-$T_c$ superconductors and it
is correctly reproduced by holographic models~\cite{Horowitz:2010gk}.
However, in the case at hand, the new phase is not superconducting,
as the neutral scalar operator preserves the $U(1)$ symmetry of the
action~\eqref{lagr_genB}. Nevertheless, also in this case we observe
a sort of inverse-Meissner effect~\cite{Albash:2008eh}: as the
external magnetic field is increased, a second order phase transition
occurs, and the condensate disappears. We stress that this effect
occurs although the scalar condensate is homogeneous on the boundary.
\subsubsection{Free energy and magnetic susceptibility of dyonic
dilatonic black branes}
\begin{center}
\begin{figure*}[ht]
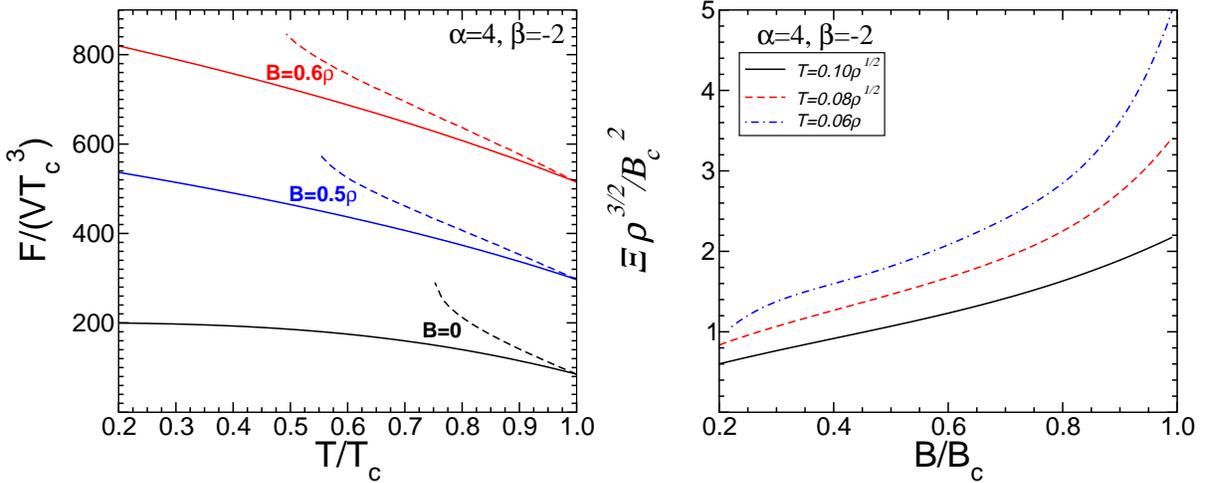

\begin{center}
\begin{tabular}{cc}
\includegraphics[scale=0.31,clip=true]{./Plots/F_VS_T.eps}&
\includegraphics[scale=0.31,clip=true]{./Plots/Chi_VS_B.eps}
\end{tabular}
\end{center}
\caption{Left panel: free energy for DDBBs (straight lines) and for
AdS-RN black holes (dashed lines) as a function of the temperature
for selected values of the magnetic field. Right panel: normalized
magnetic susceptibility for DDBBs as a function of the magnetic
field. We used $f(\phi)=\cosh(2\phi)$.
}
\label{fig:free_en}
\end{figure*}
\end{center}

As discussed above, a sufficiently strong magnetic field will destroy
the new phase. The critical magnetic field can be also understood in
terms of energetics. The difference in free energy between the normal
and the dressed phase reads
\be
\frac{B_c^2(T){\cal
V}}{8\pi}=F_\text{normal}(T)-F_\text{dressed}(T)\,,
\ee
where ${\cal V}$ is the volume of $(x,y)$ plane and $F$ is the free
energy. We wish to compare the free energy between different phases
--~with and without the scalar condensate~-- and in presence of a
magnetic field. 
As for the normal phase, the free energy of a dyonic AdS-RN black
hole (see for instance Ref.~\cite{Hartnoll:2008kx}) reads 
\be
\frac{F_\text{RN}}{{\cal V}}=\frac{F_{\text{normal}}}{{\cal V}}
=-r_h^3+\frac{3(\rho^2+B^2)}{4r_h}\,,\label{F0}
\ee
where we have set $L=1$.
In order to compute the free energy for DDBBs we start from the
Euclidean action
\be
S_E=-\int d^4x\sqrt{-G}{\cal L}\,,
\ee
where ${\cal L}$ is the Lagrangian written in~\eqref{lagr_genB} and
computed for the numerical DDBB solution.
Following Ref.~\cite{Hartnoll:2008kx} we write the Einstein tensor,
\be
E_{xx}=\frac{r^2}{2}({\cal L}-R)+\frac{B^2 f(\phi)}{2r^2}\,,
\ee
where $E_{\mu\nu}=R_{\mu\nu}-\frac{1}{2}G_{\mu\nu}R$. Thus we obtain
\be
{\cal
L}=-E^t_t-E^r_r-\frac{B^2f(\phi)}{r^4}=-\frac{1}{r^2}\left[(rg)'+(rg
e^{-\chi})'e^\chi\right]-\frac{B^2f(\phi)}{r^4}\,,
\ee
where we have used the scalar curvature $R=-E^a_a$ computed for the
dressed solution. Then the Euclidean action reads
\be
S_E=\int d^3x\int_{r_h}^\infty dr\left(2r g
e^{-\chi/2}\right)'+B^2\int d^3x\int_{r_h}^\infty
dr\frac{f(\phi)e^{-\chi/2}}{r^2}\,.
\ee
The first radial integral in the equation above is a total derivative
and it is straightforwardly performed. However, it diverges as
$r\to\infty$ and must be regularized by suitable
counterterms~\cite{Hartnoll:2008kx}. On the other hand, the
contribution arising from the magnetic field is finite. Defining the
regularized Euclidean action $\tilde{S}_E$, the thermodynamical
potential in the gran-canonical ensemble reads
\be
\Omega=T \tilde{S}_E=\int d^2x\left(\frac{-\epsilon
L^2}{2}+B^2\int_{r_h}^\infty
dr\frac{f(\phi)e^{-\chi/2}}{r^2}\right)\,,
\ee
where $\epsilon$ is the energy density, we have defined the (compact)
Euclidean time $\int dt=1/T$, and we have used the boundary
conditions (either ${\cal O}_-=0$ or ${\cal O}_+=0$) for the scalar
field. Finally, the free energy of DDBBs in the canonical ensemble
simply reads ($L=1$)
\be
F_\text{DD}\equiv  F_\text{dressed}=\Omega+\mu
Q={\cal V}\left(-\frac{\epsilon}{2}+\mu\rho+B^2\int_{r_h}^\infty
dr\frac{f(\phi)e^{-\chi/2}}{r^2}\right)\,.
\ee
In the particular case of AdS-RN black holes ($f(\phi)\equiv1$ and
$\chi\equiv0$) the integral above is trivial and $F_\text{DD}$
reduces to Eq.~\eqref{F0}.
In the left panel of Fig.~\ref{fig:free_en} we compare $F_\text{RN}$
and $F_\text{DD}$ as functions of the temperature for selected values
of the constant magnetic field. Free energies refers to solutions
with same mass and same charge. Although not shown, for both
solutions the specific heat $c=-T\partial_T F/{\cal V}$ is positive
for any
value of $B$. AdS-RN black holes always have a larger free energy,
for any value of $B$ and, roughly speaking, the magnetic field shifts
the free energy of both solutions up. Therefore, when $T<T_c$ and
$B<B_c$, DDBBs are energetically favored. However, the magnetic field
contributes to lower the difference $\Delta F=F_{RS}-F_\text{DD}$.
Such a difference grows as $T\to0$ but, for fixed temperature, it
decreases as the magnetic field increases. This is consistent with
our analytical understanding, since we expect $\Delta F=0$ both when
$T=T_c$ and when $B=B_c=\rho$.

Finally, we can compute the magnetic susceptibility
\be
\varXi=\left.{\frac{\partial^2 F}{\partial
B^2}}\right\rvert_{\rho,T}\,,\label{eq:suscept}
\ee
where the derivative is performed on solutions at constant
temperature and constant charge density. Therefore in the numerical
integration, we find the zeros of Eqs.~\eqref{root2}, in order to
compute the free energy as a function of the magnetic field at
constant temperature and we obtain the magnetic susceptibility by
performing the derivative in Eq.~\eqref{eq:suscept}. Results are
shown in the right panel of Fig.~\ref{fig:free_en}. The free energy
and the magnetic field are normalized as $F\to F/\rho^{3/2}$ and as
$B\to B/B_c$, respectively. Hence, the magnetic susceptibility is
normalized as $\varXi\to\varXi\rho^{3/2}/B_c^2$.
The magnetic susceptibility $\varXi$ is order 1 and positive. This
means that the boundary theory is strongly diamagnetic. This is
analog to the AdS-RN case where $\varXi\sim9/(8\pi T)>0$ for $B\ll M$
and $\rho\ll M$~\cite{Hartnoll:2008kx}.
\subsection{Holographic properties of DDBBs}
Let us now discuss some holographic properties of DDBBs. In
particular we shall focus on  the effects of the magnetic field on the
electrical conductivity in the dual theory, such as the Hall effect
and the presence of cyclotron resonances. The AdS/CFT correspondence
provides a
precise prescription for the computation of electrical conductivity 
of the dual field theory in terms of
bulk electromagnetic perturbations. However, electromagnetic
perturbations of a dyonic black hole are fairly involved. In fact,
 the
minimal set of perturbations includes both the  $x$ and $y$ 
components of the gauge potential $A_x(r)$, $A_y(r)$ and the $tx$ and 
$ty$ components of the metric, $G_{tx}(r)$
and $G_{ty}(r)$, which are coupled through the electric and magnetic
field. 

Perturbations of dyonic AdS-RN black holes have been studied
in Refs.~\cite{Hartnoll:2007ai,Hartnoll:2007ip}. Here we want to
extend those calculations to the case of dilatonic background. 

Let us consider perturbations with vanishing $3$-momentum, 
\be
A_\mu=(A_0,0,0,Bx)+(0,0,A_x(r),A_y(r))e^{-i\omega t}\nn\,,
\ee
and similarly for the metric perturbations, $G_{tx}(r)e^{-i\omega t}$
and $G_{ty}(r)e^{-i\omega t}$. 
Linearized Einstein and Maxwell equations provide a set of four
coupled equations. Two of them are
\beq
A_x''+A_x'\left(\frac{f'(\phi)}{f(\phi)}\phi'+\frac{g'}{g}-\frac{\chi'}{2}\right)+\omega^2\frac{e^\chi}{g^2}A_x&=&-e^\chi\left[\frac{iB\omega}{r^2g^2}G_{ty}+\frac{A_0'}{g}\left(G_{tx}'-\frac{2}{r}G_{tx}\right)\right],\label{EM_pert1}\\
G_{tx}'-\frac{2}{r}G_{tx}+f(\phi)A_0' A_x&=&-\frac{iB
f(\phi)}{r^2\omega}\left[A_0' G_{ty}+g e^{-\chi}
A_y'\right]\,,\label{EM_pert2}
\eeq
and the other two can be obtained from those above by changing
$x\leftrightarrow y$ and $B\leftrightarrow-B$.
Notice that terms proportional to $B$ couple perturbations along the
$x$ direction to those along the $y$ direction. When $B=0$ equations
above decouple and reduce to a single Schr\"odinger-like equation. In
the case at hand, such a decoupling does not occur and we are left
with a system of four coupled ODEs. Furthermore notice the presence
of terms proportional to $B/\omega$. When $B\neq0$ these terms
diverges in the $\omega\to0$ limit, whereas they vanish if $B=0$.
Thus, as first noted in Ref.~\cite{Hartnoll:2007ai}, the limits
$B\to0$ and $\omega\to0$ do not commute.

We integrate the system of ODEs above numerically, starting from a
series expansion close to the horizon, where we impose purely ingoing
waves. The asymptotical behaviors read
\beq
A_x&\sim& a_x^{(0)}+ a_x g_h^\nu\,,\qquad A_y\sim a_y^{(0)}+ a_y
g_h^\nu\,,\\
G_{tx}&\sim& g_x^{(0)}+g_x g_h^{\nu+1}\,,\qquad G_{ty}\sim
g_y^{(0)}+g_y g_h^{\nu+1}\,,
\eeq
with $g_h=g(r_h)\sim(r-r_h)$ and the requirement of purely ingoing
waves at the horizon implies $\nu=-i\omega e^{\chi_h/2}/g'(r_h)$. The
constants $a_x$, $a_x^{(0)}$, $g_x$, $g_x^{(0)}$, $a_y$, $a_y^{(0)}$,
$g_y$ and $g_y^{(0)}$ are related to each other by requiring that the
expansions above are solutions of the equations of motion at first
order.
\subsubsection{Conductivity in the dual field theory}
Before presenting the results of the numerical integration, we briefly
review some analytical results obtained in
Refs.~\cite{Hartnoll:2007ai,Hartnoll:2007ip} for the electrical
conductivity in theories dual to AdS-RN  dyonic black hole. In that
case,
Eqs.~(\ref{EM_pert1})-(\ref{EM_pert2}) can be solved analytically in
the hydrodynamical limit, i.e. when $\omega/T\ll\mu/T,B/T^2$. In this
limit the diagonal and off-diagonal components
of the conductivity matrix ${\bf \sigma}$ can be computed via the
AdS/CFT duality and they
read~\cite{Hartnoll:2007ip}
\begin{equation*}
\sigma_{xx} =
\sigma_Q \frac{ \omega (\omega + i \gamma + i
\omega_c^2/\gamma)}{(\omega+i \gamma)^2 - \omega_c^2} \,,\qquad
\sigma_{xy} = -\frac{\rho}{B} \frac{-2 i \gamma \omega + \gamma^2 +
\omega_c^2}{(\omega+ i \gamma)^2 - \omega_c^2} \ ,
\end{equation*}
with
\be
\label{cyclotronpole}
\omega_c = \frac{B \rho}{\epsilon+{\cal P}}\,,\quad \gamma =
\frac{\sigma_Q B^2}{\epsilon+{\cal P}} \,,\quad\sigma_Q =
\frac{(sT)^2}{(\epsilon+{\cal P})^2} 
\ee
and where ${\cal P}$, $s$ and $\epsilon$ are the pressure, entropy
density and the energy
density respectively. Rotational invariance implies
$\sigma_{xx}=\sigma_{yy}$ and $\sigma_{xy}=-\sigma_{yx}$. The
electrical conductivity has a pole at $\omega = \omega_c - i \gamma$,
corresponding to a damped cyclotron frequency. 
The real part $\omega_c$ does not depend on the
temperature~\cite{Hartnoll:2007ip}. 
Furthermore, in the $\omega\to0$ limit, the DC diagonal component
$\sigma_{xx}$ vanishes, whereas the DC off-diagonal component gives
the well-known Hall conductivity: 
\be
\sigma_{xx}=0\,,\qquad \sigma_{xy}=\frac{\rho}{B}\,.
\ee
Notice that $\sigma_{xy}\to\infty$ as $B\to0$. This is due to the
non-commuting limits $\omega\to0$ and $B\to0$.

It is important to study whether and how the scalar condensate affects
these results. Unfortunately, in our model, the background solution
is only known numerically and this prevents to derive explicit
formulas. However we can still use the AdS/CFT prescription in order
to relate the conductivity to the asymptotic behavior of the
numerical solution. Following Ref.~\cite{Hartnoll:2007ip}, the
conductivity reads
\be
\sigma_\pm=\sigma_{xy}\pm i\sigma_{xx}=\frac{{\cal B}_x\pm i {\cal
B}_y}{{\cal E}_x\pm i {\cal E}_y}\,,\label{sigmapm}
\ee
where
\be
{\cal B}_i=-\lim_{r\to\infty}\epsilon_{ij}A_j'\,,\qquad {\cal
E}_i=\lim_{r\to\infty}\left[f(\phi)\left(i\omega
A_i-\frac{B}{r^2}\epsilon_{ij}G_{tj}\right)\right]\,,\label{EandB}
\ee
are the spatial components of $\delta F$ and $\delta \star F$
respectively ($F=F_0+\delta F$ and $\star F=\star F_0+\delta\star F$
being its dual). The coefficients $A_i^{(0)}$, $A_i^{(1)}$ and
$G_{ti}^{(0)}$ are related to the asymptotic behavior of the
electromagnetic and metric perturbations at infinity:
\be
A_i\to A_i^{(0)}+\frac{A_i^{(1)}}{r}\,,\qquad G_{ti}\to G_{ti}^{(0)}
r^2\,,\quad i=x,y
\ee
Thus, once perturbation equations are solved with suitable boundary
conditions, Eq.~\eqref{sigmapm} gives the AdS/CFT prescription for
the conductivity in the dual theory.
\subsubsection{Numerical results}
\begin{center}
\begin{figure*}[ht]
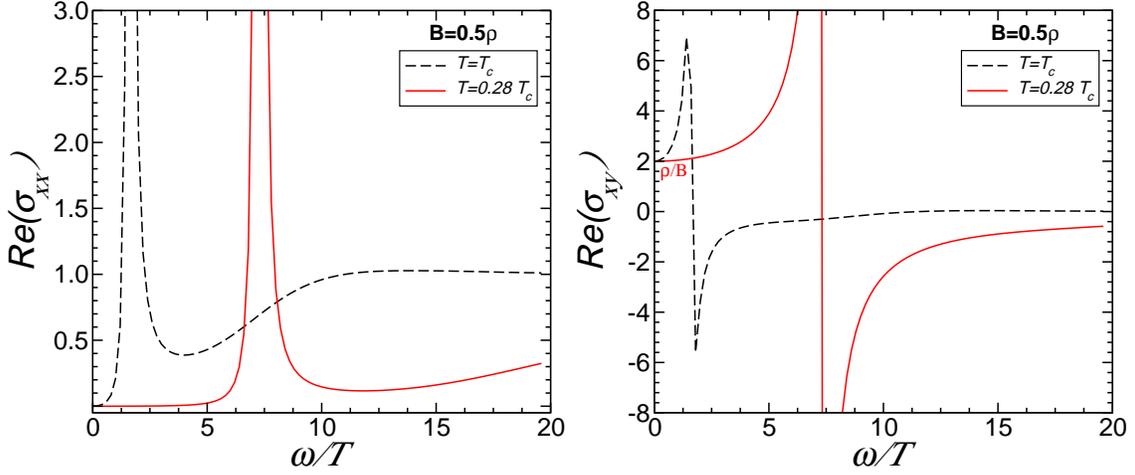

\begin{center}
\begin{tabular}{cc}
\includegraphics[scale=0.31,clip=true]{./Plots/sigmaxx_B05.eps}&
\includegraphics[scale=0.31,clip=true]{./Plots/sigmaxy_B05.eps}
\end{tabular}
\end{center}
\caption{Conductivity in the field theory  dual DDBBs  with
$f(\phi)=\cosh(2\phi)$ (left panel: $\sigma_{xx}$,
right panel: $\sigma_{xy}$) as a function of the frequency for
$B/\rho=0.5$. The conductivity in the
normal phase at $T=T_c$ (AdS-RN black brane) is compared to that in
the dressed phase at $T\sim0.28 T_c$ (DDBB).}
\label{fig:conductivity}
\end{figure*}
\end{center}

\begin{center}
\begin{figure*}[ht]
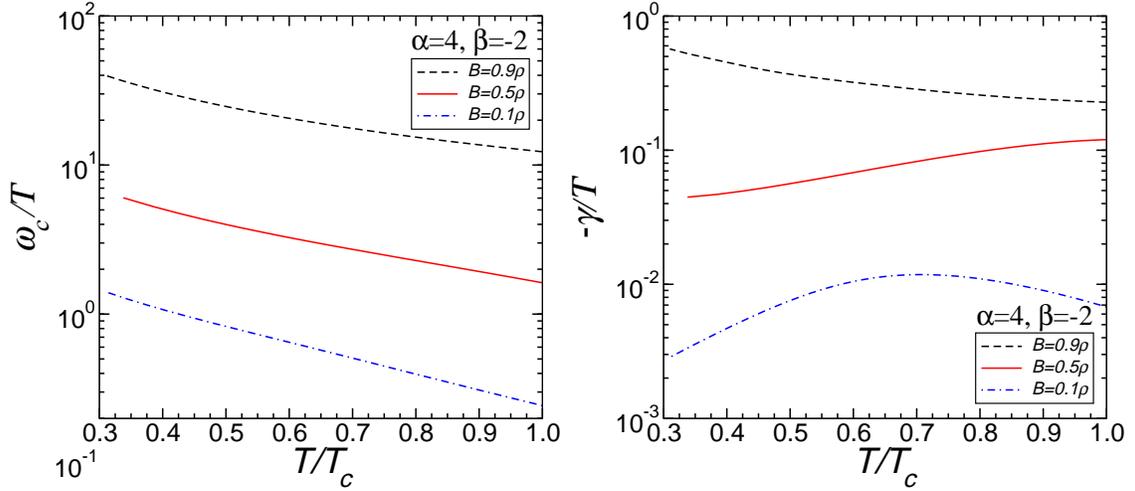

\begin{center}
\begin{tabular}{cc}
\includegraphics[scale=0.31,clip=true]{./Plots/poleR_a2.eps}&
\includegraphics[scale=0.31,clip=true]{./Plots/poleI_a2.eps}
\end{tabular}
\end{center}
\caption{Real (left panel) and imaginary (right panel) part of the
damped cyclotron frequency $\omega=\omega_c+i\gamma$ as functions of
the temperature for selected values of the magnetic field $B/\rho$ 
for DDBBs with $f(\phi)=\cosh(2\phi)$.}
\label{fig:cyclotron}
\end{figure*}
\end{center}

In Fig.~\ref{fig:conductivity}  we show the conductivities
$\sigma_{xx}$ and
$\sigma_{xy}$ as functions of the frequency  both for the
AdS-RN case ($T=T_c$) and for the DDBB at $T<T_c$. The numerical
procedure previously discussed has been tested by reproducing
numerical results in Ref.~\cite{Hartnoll:2007ip} for vanishing scalar
field. A general result that can be inferred from our simulations is
that, regardless of  the scalar condensate, the DC conductivities are
the
same as those computed for AdS-RN black
branes~\cite{Hartnoll:2007ip}, 
\be
\sigma_{xx}(\omega\to0)=0\,,\qquad
\sigma_{xy}(\omega\to0)=\frac{\rho}{B}\,,
\ee
at any $T\leq T_c$. This result holds regardless of  the precise form
of
$V(\phi)$ and $f(\phi)$ given by~\eqref{expVandfB}.
While the first result ($\sigma_{xx}=0$) simply arises from the
Lorentz invariance, it is interesting that the Hall effect is not
affected by the scalar condensate, for any value of $B$.
However, as shown in Fig.~\ref{fig:conductivity}, the AC behavior is
more complex. Depending on the temperature and on the magnetic field,
sharp peaks appear in the real part of $\sigma_{xx}$ and
$\sigma_{xy}$. These correspond to the (complex) cyclotron
frequencies discussed in Ref.~\cite{Hartnoll:2007ip} for AdS-RN black
branes. Our results confirm and extend that analysis to the case of
DDBBs. Indeed the scalar condensate affects the cyclotron frequency.
In Fig.~\ref{fig:cyclotron} we show the location of the pole of
$\sigma_{xx}$ as a function of the temperature for selected values of
$B$. Both the real and the imaginary part of the frequency strongly
depend on the magnetic field (notice that the plot scale in
Fig.~\ref{fig:cyclotron} is logarithmic). The real part increases
exponentially as the temperature is lowered, while the imaginary
part, as a function of the temperature, has a less clear behavior,
being monotonic at large values of $B$ and having a non-monotonic
behavior at small values of $B$. 
%
%

%

\section{Conclusion}\label{conclusion}
In this paper we have presented a detailed  study 
of the holographic properties of 
the 4D, charged, black brane solutions  of   broad 
classes of EMDG models both at finite and vanishing temperature and
for 
different charge configurations (purely electric or magnetic, dyonic).
Although our 
analysis  is far from being  exhaustive, it is likely that
most of the qualitative relevant features  concerning the 
holographic behavior of 4D charged dilatonic black branes in EMDG 
models  
have been captured by our investigation.

For what concerns the zero temperature limit, our results extend 
the results of Ref.
\cite{Goldstein:2009cv,Cadoni:2009xm,Goldstein:2010aw} 
obtained in the case of a exponential coupling function $f$ 
to a broad class of EMDG models.
In the $T\sim 0$ regime the AC conductivity has the universal scaling 
behavior, $\sigma_{AC}\sim\omega^{2}$ as a
function 
of the frequency $\omega$. For a generic form of the self-interaction 
potential $V(\phi)$ the quadratic behavior is generalized to a 
generic  power-law form $\sigma_{AC}\sim\omega^{s}$. 

In view of the results of Ref.
\cite{Goldstein:2009cv,Cadoni:2009xm,Goldstein:2010aw}
one could have been lead to conclude that 
the power-law behavior of the conductivity is related to the 
Lifshitz form of the black brane metric in the near-extremal, 
near-horizon region. However, we have shown that this power-law 
behavior of $\sigma_{AC}$ is not spoiled if the Lifshitz form of 
the metric is (multiplicatively) deformed by  powers of logarithms.
This means that a power-law scaling behavior of $\sigma_{AC}$ is 
still consistent with a breaking of the non-relativistic scaling  
isometry of the Lifshitz background.

On the other hand the behavior of the DC conductivity in the zero 
temperature limit  as a function of $T$ seems more involved. 
The quadratic  scaling $\sigma_{DC}\sim T^{2}$ found in Ref.
\cite{Goldstein:2010aw} 
 for the case of a exponential coupling function $f$, 
seems to be a quite general feature of EMDG models. 
This gives further support to the `charged plasma picture' proposed 
in Ref. \cite{Goldstein:2010aw}. The electric conductivity is 
suppressed at small temperatures by strong repulsion. 
However, differently from the case of  $\sigma_{AC}(\omega)$
discussed above, now 
the deformation of the Lifshitz form of the extremal black brane
metric 
by  logarithms, due to a power-law self-interaction potential,
changes  $\sigma_{DC}(T)$ (see Eq. 
\eqref{k1}). This different behavior of $\sigma_{AC}(\omega)$ and
$\sigma_{DC}(T)$ in the near-extremal case is somehow puzzling and 
a proper understanding of it could shed light on the physical nature 
of the dual field theory.

For what concerns black branes at finite temperature our results
extend 
the results of Ref. \cite{Cadoni:2009xm}, derived 
for the case of   coupling  functions $f\sim\cosh\alpha 
\phi$ and $f\sim1+\alpha 
\phi^{2}$, to a broad class of EMDG models. The non-monotonic
behavior 
of $\sigma_{AC}(\omega)$ --~characterized by a
minimum at low frequencies and then by a `Drude peak', 
at $\omega= 0$~-- and of
$\sigma_{DC}(T)$ --~characterized by the presence of a maximum,
reminiscent 
of the Kondo effect~--  seem to be a rather generic feature of EMDG.
In particular, we have shown that the emergence of these effects is 
not related to  the presence of a phase transition RN/CDBB (as it is 
the case for the models of Ref.  \cite{Cadoni:2009xm}) but  shows up
because of the nonmimimal coupling, whenever a sufficiently large
scalar condensate  is 
present.  

Finally, our results concerning the holography of dyonic black branes 
at finite temperature confirm and extend to a broad class of EMDG 
the results of Ref. \cite{Goldstein:2010aw}, derived  for an
exponential 
coupling function $f$. Again, we have found that the main features 
observed  in Ref. 
\cite{Goldstein:2010aw} (Hall effect, presence of synchrotron
resonances) 
when a magnetic field is switched on,
also apply to a the case of a coupling function $f$ that allows for a 
RN/CDBB phase transition. On the other hand, the switching on of a 
magnetic field does not shed much light on the nature of the
microscopic 
degrees of freedom  of the  field theory dual to the models 
investigated in Ref. \cite{Cadoni:2009xm}. Naively, one could have 
expected this to be the case in view of the peculiar behavior of the 
AC and DC conductivities in absence of a magnetic field.
Unfortunately, the effects of the magnetic 
field on the conductivities are so strong that any other effect
becomes 
subleading and is 
completely washed out.

The overall  picture emerging from our results about the holographic 
features of charged dilatonic black branes in EDGM can be described  
as an
interpolation between some aspects reminiscent of electron motion in
real metals at finite temperature and a charged
plasma at zero temperature. It would be
highly desirable to develop analytical methods to give further
support to this picture.

\begin{acknowledgements}
This work was partially supported by FCT - Portugal through projects
PTDC/FIS/098025/2008, PTDC/FIS/098032/2008 and by the {\it
DyBHo--256667} ERC Starting
Grant.
\end{acknowledgements}


\appendix
\section{Purely magnetic dilatonic black branes via the
electromagnetic duality}\label{appendix}
In this appendix, we focus on the purely magnetic solutions, hereafter
setting $Q=0$. As discussed in the main text, in this case the AdS-RN
black
brane is stable and no magnetically charged dilatonic black branes
exist. However, we can obtain purely magnetic solutions by applying
the electromagnetic duality. In fact, a background solution
characterized by coupling function, electric charge and magnetic
charge $\{f(\phi),Q,B\}$ is related, via the electromagnetic duality,
to the same background solution with $\{h=1/f,B,-Q\}$. In particular
the charged dilatonic black branes found in Ref.~\cite{Cadoni:2009xm}
are also valid magnetic solutions with $f\to1/f$ and $B\to -Q$.

For dyonic AdS-RN black branes, the duality acts on the
electromagnetic tensor only 

\be
\frac{2\pi}{g^2}F\rightarrow \star
F\equiv\frac{\sqrt{-g}}{4}\epsilon_{\mu\nu\rho\sigma}F^{\rho\sigma}dx^\mu
\wedge dx^\nu\,.
\ee
Perturbation equations for $\delta F$ ($F=F_0+\delta F$) are most
conveniently written in terms of ${\cal B}_a$ and ${\cal E}_a$, (cf.
Eq.~\eqref{EandB}) which are the spatial component of $\delta F$ and
of $\delta \star F$ respectively. The electromagnetic duality acts as
${\cal E}\to {\cal B}$ and ${\cal B}\to - {\cal E}$. It follows from
Eq.~\eqref{sigmapm} that the duality transforms the conductivities as
\be
\sigma_\pm(Q,B)\rightarrow
\frac{1}{\sigma_\pm(B,-Q)}\,.\label{cond_RN}
\ee
Equation~\eqref{cond_RN} has been recently extended to the case of
SL(2,R) invariant theories with a dilaton and an
axion~\cite{Goldstein:2010aw}. Also in that case, starting from a
purely electrical background, a transformation similar
to~\eqref{cond_RN} holds
\be
\sigma_\pm(0,-Q)=\frac{1}{\sigma_\pm(Q,0)}\,,\label{eq:SL2R}
\ee
which does not explicitly depend on the dilaton and axion fields.
Although Eq.~\eqref{eq:SL2R} has been derived for SL(2,R) invariant
theories and the action~\eqref{lagr_genB} is not invariant under
the full SL(2,R) transformation, nevertheless the same derivation
should apply to our case
as well. We shall explicitly confirm this statement below, by
computing the conductivity in the purely magnetic case. 
\subsubsection*{Electrical conductivity}
Notice that, in the rest of this section, we keep referring to the
coupling $f(\phi)$ for convenience, although the duality
transformation acts on the action~\eqref{lagr_genB} by transforming
$f\to h=1/f$. The real non-minimal coupling for solutions obtained by
the duality is $h(\phi)$, not $f(\phi)$. Furthermore, note that
$h(\phi)\sim1-\alpha\phi^2$ at $\phi\sim0$, whereas
$f(\phi)\sim1+\alpha\phi^2$. The electromagnetic duality reverses the
sign of $\alpha$. This is why purely magnetic solutions exist in this
case (cf. Eq.~\eqref{eff_massB}).

For purely magnetic backgrounds, setting $A_0(r)\equiv0$,
Eqs.~\eqref{EM_pert1}-\eqref{EM_pert2} (and those obtained from them
by $x\leftrightarrow y$ and $B\leftrightarrow-B$) decouple pairwise.
Equations for $A_x$ and $G_{ty}$ read
\beq
A_x''+A_x'\left(\frac{f'(\phi)}{f(\phi)}\phi'+\frac{g'}{g}-\frac{\chi'}
{2}\right)+\omega^2\frac{e^\chi}{g^2}A_x+\frac{iB\omega}{r^2g^2}e^\chi
G_{ty}&=&0\,,\label{EM_pert1B}\\
G_{ty}'-\frac{2}{r}G_{ty}-\frac{iB f(\phi)}{r^2\omega}g e^{-\chi}
A_x'&=&0\,,\label{EM_pert2B}
\eeq
and those for $A_y$ and $G_{tx}$ can be again obtained by
$x\leftrightarrow y$ and $B\leftrightarrow -B$. 
The two equations above can be written in terms of a single
Schr\"odinger-like
equation for $A_x'(r)$, \text{i.e.} they are third order in $A_x(r)$.
In fact, we define
\be\label{a1}
A_x'(r)=\frac{e^{\chi/2}}{g(r)\sqrt{f(\phi)}}Y(r)\,,
\ee
and the equation for $Y(r)$ reads
\be\label{a2}
Y''(r)+\left(\frac{g'}{g}-\frac{\chi'}{2}\right)Y'(r)+
\frac{e^\chi}{g^2}\left[\omega^2-V_{s}¥(r)\right]Y(r)=0\,,
\ee
or, equivalently, 
\be\label{a3}
Y''(z)+\left[\omega^2-V_{s}¥(z)\right]Y(z)=0\,,
\ee
where $z$ is the tortoise coordinate defined by
$dr/dz=e^{\chi(r)/2}/g(r)$ and the explicit form of the potential
reads
\be
V_{s}¥(z)=ge^{-\chi}\left\{\frac{B^2f(\phi)}{r^4}-
\frac{g}{2}\frac{f'(\phi)}{f(\phi)}\left[\phi''+
\left(\frac{f''(\phi)}{f'(\phi)}-\frac{3}{2}
\frac{f'(\phi)}{f(\phi)}\right)\phi'^2+
\left(\frac{g'}{g}-\frac{\chi'}{2}\right)\phi'\right]\right\}\,.\label{pot_magnetic}
\ee
Interestingly, in the equation above the magnetic field dependence is
$B^2$ and the contribution $\propto 1/\omega$ of
Eq.~\eqref{EM_pert2B} cancels out, i.e. in the purely magnetic case
the limits $B\to0$ and $\omega\to0$ commute.

\begin{center}
\begin{figure*}[ht]
\begin{center}
\begin{tabular}{cc}
\includegraphics[scale=0.31,clip=true]{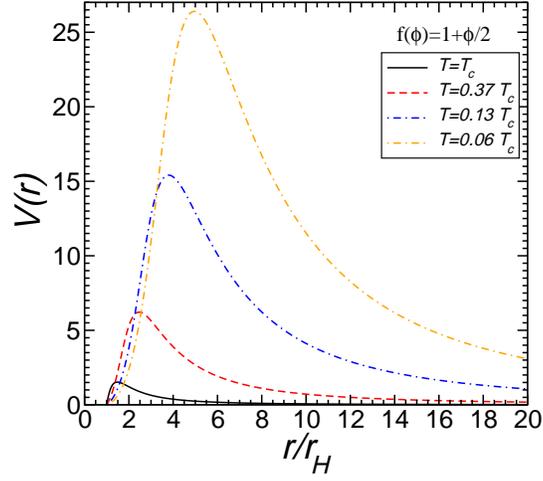}
\end{tabular}
\end{center}
\caption{Schr\"odinger potential for electric perturbations around
purely magnetic dilatonic black branes  (DBBs) (cf.
Eq.~\eqref{pot_magnetic}  with $f={1}/{h}=1+\phi/2$.}
\label{fig:pot_magnetic}
\end{figure*}
\end{center}

For a magnetic AdS-RN black brane ($\chi\equiv0$ and $\phi\equiv0$)
the potential above simply reduces to
\be
V_{RS}=g_{RS}(r)\frac{B^2}{r^4}\,,
\ee
and it is positive defined. Moreover $V_{RS}=0$ at the horizon and at
infinity. From general quantum mechanics theorems it follows that
such a Schr\"odinger potential does not admit bound states, i.e. the
magnetic AdS-RN black brane is stable. The potential for purely
magnetic dilatonic solutions (obtained via the electromagnetic
duality) is shown in Fig.~\ref{fig:pot_magnetic}. Also in this case
the potential is positive defined and the background solution is
stable.
%
\subsubsection*{Numerical results}
\begin{center}
\begin{figure*}[ht]
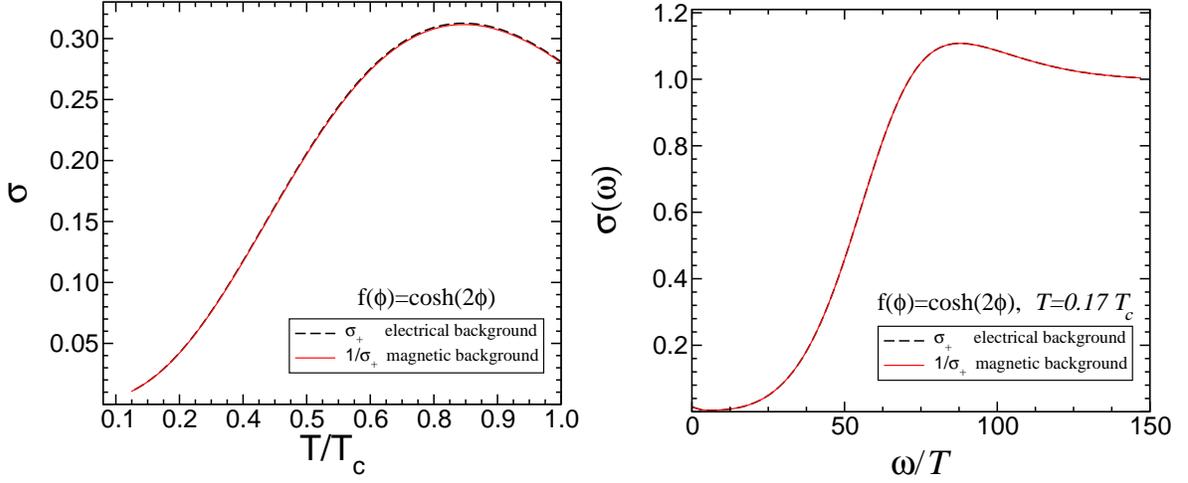

\begin{center}
\begin{tabular}{cc}
\includegraphics[scale=0.31,clip=true]{./Plots/sigmap_VS_T.eps}&
\includegraphics[scale=0.31,clip=true]{./Plots/sigmap_VS_omega_EoM.eps}
\end{tabular}
\end{center}
\caption{Left panel: comparison between the DC conductivity in an
electrical and in a magnetic background. The black dashed line is
$\text{Re}[\sigma_+(\omega=0)]$ in the purely electrical background,
whereas the red straight line is
$\text{Re}[{\sigma_+}^{-1}(\omega=0)]$ in the purely magnetic
background with $f(\phi)\to1/f(\phi)$. Right panel: the same but
showing the AC conductivity at $T=0.17 T_c$ as a function of the
frequency. As proved in Ref.~\cite{Goldstein:2010aw}, the two
functions coincide. The magnetic solution is obtained via the
electromagnetic duality from an electrical solution with
$f(\phi)=\cosh(\sqrt{\alpha}\phi)$.}
\label{fig:purely_magnetic}
\end{figure*}
\end{center}

\begin{center}
\begin{figure*}[ht]
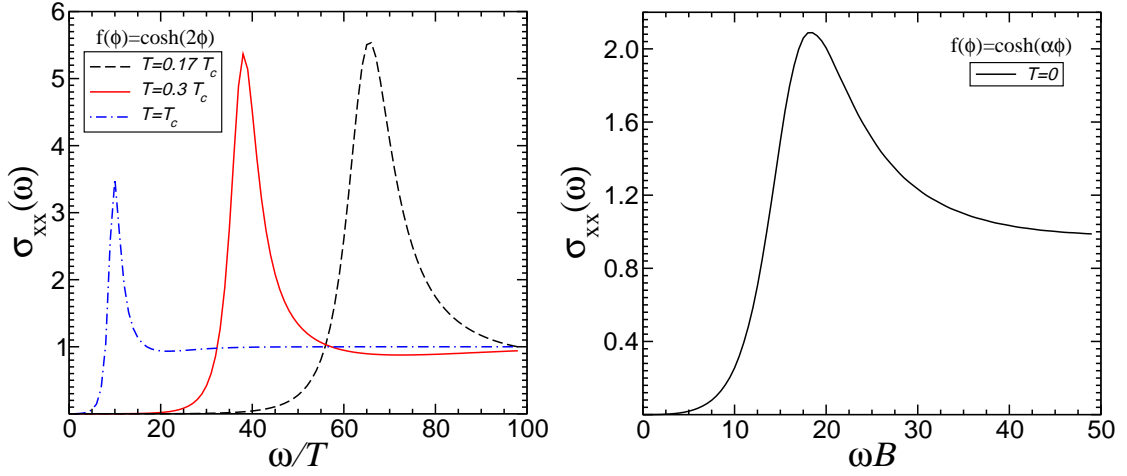

\begin{center}
\begin{tabular}{cc}
\includegraphics[scale=0.31,clip=true]{./Plots/sigmaxx_VS_omega_magnetic.eps}&
\includegraphics[scale=0.31,clip=true]{./Plots/sigmaxx_VS_omega_magnetic_T0.eps}
\end{tabular}
\end{center}
\caption{Left: AC conductivity in a purely magnetic background for
several values of $T/T_c$ and for $f(\phi)=\cosh(2\phi)$. Right: AC
conductivity $\sigma_{xx}$ for a purely magnetic DBB
at $T=0$. The magnetic solution is obtained via the electromagnetic
duality from the purely electric DBB with
$f(\phi)=\cosh(\phi/\sqrt{3})$ (see Ref.~\cite{Cadoni:2009xm} for
details). For purely magnetic backgrounds the off-diagonal component
identically vanishes, i.e. $\sigma_{xy}(\omega)\equiv0$.}
\label{fig:purely_magnetic_xx}
\end{figure*}
\end{center}

As non-trivial test of our numerical method, we have computed the
conductivity $\sigma_+(\omega)$  in a purely
electrical background and compared it to the inverse of the
conductivity in a purely magnetic background $\sigma_+^{-1}(\omega)$
with $f(\phi)\to1/f(\phi)$. A representative example is shown in
Fig.~\ref{fig:purely_magnetic}. The two functions coincide, confirming
analytical expectations (cf. Eq.~\eqref{eq:SL2R} and
Ref.~\cite{Goldstein:2010aw}).
Notice that, although the electromagnetic duality straightforwardly
relates the conductivities $\sigma_\pm$ in an electrical and in a
magnetic background, nevertheless the transformation is non-trivial
and it connects the conductivities computed in two different theories
(the couplings in the action are different). Furthermore, not only
$\sigma_\pm$ are complex quantities, but, above all, the physically
interesting quantities are the frequencies $\sigma_{xx}$ and
$\sigma_{xy}$. Namely, the explicit transformations for these
quantities read
\be
\sigma_{xy}^M=\frac{\sigma_{xy}^E}{\left(\sigma_{xx}^E\right)^2+\left(\sigma_{xy}^E\right)^2}\,,\quad
\sigma_{xx}^M=-\frac{\sigma_{xx}^E}{\left(\sigma_{xx}^E\right)^2+\left(\sigma_{xy}^E\right)^2}\,,\quad
\text{Re}[\sigma_M]=\frac{\text{Re}[\sigma_E]}{\text{Abs}[\sigma_E]^2}\,,\quad
\text{Im}[\sigma_M]=-\frac{\text{Im}[\sigma_E]}{\text{Abs}[\sigma_E]^2}\,,
\ee
where $\sigma^M$ and $\sigma^E$ are the electrical conductivities in
the magnetic and electrical case, respectively and all the quantities
are complex. Therefore the explicit dependence of, say,
$\text{Re}\left[\sigma^M_{xx}\right]$ can be non-trivial. In the left
panel of Fig.~\ref{fig:purely_magnetic_xx} we show the AC
conductivity $\sigma_{xx}(\omega)$ for a purely magnetic background
solution. From our numerical simulations we can infer the general
behavior
\be
\sigma_{xy}(\omega)\equiv0\,,\qquad \sigma_{xx}(\omega\sim0)=0\,,
\ee
that is, for any temperature, the off-diagonal component of the
conductivity vanishes at any frequency. However, as shown in the left
panel of Fig.~\ref{fig:purely_magnetic_xx}, the AC diagonal component
has a maximum whose location and height depend on the temperature.
These peaks in the conductivity may signal the excitation of some
bound state in the dual field theory. Thus $\sigma^M$, although
related to $\sigma^E$ by the electromagnetic duality, can show some
non-trivial features.
%
\subsubsection*{The Zero temperature limit of purely magnetic DDBs}
For the sake of completeness, let us conclude this section by briefly
discussing extremal magnetic DBBs. Purely magnetic background
solutions can be again obtained from
purely electrical extremal backgrounds via the electromagnetic
duality. Extremal electrical solutions have been discussed in detail
in Ref.~\cite{Cadoni:2009xm}. They are most
conveniently studied using the ansatz~\eqref{ansatz_T0}.
Using this ansatz perturbation equations around dyonic background
solutions read
\beq
A_x''+A_x'\left(\frac{f'(\phi)}{f(\phi)}\phi'+\frac{\lambda'}{\lambda}\right)+\frac{\omega^2}{\lambda^2}A_x+\frac{iB\omega}{H^2\lambda^2}G_{ty}+\frac{A_0'}{\lambda}\left(G_{tx}'-\frac{2H'}{H}G_{tx}\right)&=&0\,,\label{EM_pert1T0}\\
G_{tx}'-\frac{2H'}{H}G_{tx}+f(\phi)A_0' A_x+\frac{iB
f(\phi)}{H^2\omega}\left[A_0' G_{ty}+\lambda
A_y'\right]&=&0\,,\label{EM_pert2T0}
\eeq
plus those obtained via $x\leftrightarrow y$ and
$B\leftrightarrow-B$. As previously discussed, in a purely magnetic
background ($A_0(r)\equiv0$) perturbation equations decouple and can
be written as a Schr\"odinger equation. Following the derivation
described by Eqs. (\ref{a1})-(\ref{pot_magnetic}), we obtain a
Schr\"odinger-like
equation, $Y''(z)+\left[\omega^2-V_{s}¥(z)\right]Y(z)=0$, where the
potential reads
%
\be
V_{s}¥(z)=\lambda\left\{\frac{B^2f(\phi)}{H^4}-\frac{\lambda}{2}
\frac{f'(\phi)}{f(\phi)}\left[\phi''+\left(\frac{f''(\phi)}{f'(\phi)}-
\frac{3}{2}\frac{f'(\phi)}{f(\phi)}\right)\phi'^2+\frac{g'}{g}\phi'
\right]\right\}\,.\label{pot_magneticT0}
\ee
where $dr/dz=1/\lambda$ and $Y(z)=\sqrt{f(\phi)}\lambda A'_{x}¥$.
The conductivity as a function of the frequency at $T=0$ is shown in
the right panel of Fig.~\ref{fig:purely_magnetic_xx}. Also in this
case the real part of $\sigma_{xx}(\omega)$ shows a peak at
intermediate frequencies.

\bibliography{CDBHs2}
\end{document}